\documentclass[pdflatex,sn-mathphys-num]{sn-jnl}

\usepackage{amsmath,amssymb,amsfonts}
\usepackage{amsthm}
\usepackage{mathrsfs}
\usepackage{mathtools}
\usepackage{bbm}
\usepackage{dutchcal}
\usepackage{nicefrac}

\usepackage{graphicx}
\usepackage{epsfig}
\usepackage{float}

\usepackage[table]{xcolor}
\usepackage{array}
\usepackage{multirow}
\usepackage{booktabs}
\usepackage{hhline}
\usepackage{siunitx}

\usepackage{caption}
\usepackage{subcaption}

\usepackage{algorithm}
\usepackage{algorithmicx}
\usepackage{algpseudocode}
\usepackage{listings}

\usepackage{enumerate}
\usepackage[inline]{enumitem}

\usepackage[title]{appendix}
\usepackage{textcomp}
\usepackage{manyfoot}
\usepackage{latexsym}
\usepackage{varioref}
\usepackage{lineno}
\usepackage{mdframed}
\usepackage[utf8]{inputenc}
\usepackage{url}

\newenvironment{rcodeboxtable}[2]
{%
  \refstepcounter{table}%
  \begin{mdframed}[
    nobreak=true,
    hidealllines=true,
    innerleftmargin=5pt,
    innerrightmargin=5pt,
    innertopmargin=0pt,
    innerbottommargin=0pt,
    skipabove=6pt,
    skipbelow=6pt
  ]%
  \noindent\textbf{Table \thetable.} #1\label{#2}

  \vspace{4pt}
}
{%
  \end{mdframed}
}

\newcommand{\beq}{\begin{equation}}
\newcommand{\eeq}{\end{equation}}

\newcommand{\cond}{\xrightarrow{d}}

\DeclarePairedDelimiter{\floor}{\lfloor}{\rfloor}

\theoremstyle{thmstyleone}%
\newtheorem{theorem}{Theorem}
\newtheorem{prop}{Proposition}
\newtheorem{lemma}{Lemma}

\theoremstyle{thmstyletwo}%

\theoremstyle{thmstylethree}%
\newtheorem{definition}{Definition}%

\begin{document}

\title[TENT]{Turing--Entropic Tail Classification: A Nonparametric Approach to Tail Inference}


\author*[1]{\fnm{Jialin} \sur{Zhang}}\email{jzhang@math.msstate.edu}

\author[2]{\fnm{Zhiyi} \sur{Zhang}}\email{zzhang@charlotte.edu}

\affil*[1]{\orgdiv{Department of Mathematics and Statistics}, \orgname{Mississippi State University}, 
\orgaddress{\city{Mississippi State}, \state{MS 39762}, \country{USA}}}

\affil[2]{\orgdiv{Department of Mathematics and Statistics}, \orgname{University of North Carolina at Charlotte}, 
\orgaddress{\city{Charlotte}, \state{NC 28223}, \country{USA}}}


\abstract{
This article introduces the Turing--Entropic Tail Classifier (TENT), a nonparametric information-theoretic framework for tail inference in discrete and continuous distributions. Central to the approach is the tail profile, a collection of information-theoretic quantities motivated by Turing's formula and domain-of-attraction theory on countable alphabets. TENT classifies the qualitative form of tail decay by comparing empirical tail profiles with theoretical benchmarks corresponding to exponential-type, near-exponential, sub-exponential, and power-law regimes, including Zipf and Pareto-type behavior. For heavier-than-exponential tails, the framework further yields point and interval estimates for selected tail parameters. Although the method is primarily designed to support preliminary model selection for discrete parametric families, a supporting result shows that, under suitable regularity conditions, discretizing continuous observations by common-width binning preserves the relevant tail decay rate. This provides a principled route for extending the classifier to continuous data. Simulation studies illustrate the finite-sample performance of TENT across multiple tail classes and sampling regimes.
}




\maketitle

\section{Introduction}

``In certain situations it is of interest to draw inference about the behavior of a distribution function in the tails without assuming that a particular parametric form for the distribution function holds globally'' \cite{hill1975simple}. This task is inherently complex for two primary reasons. First, the distributional behavior in the tail may differ markedly from its behavior over the rest of the sample space. Consequently, it is not always clear how observations from the high-probability region of the sample space should inform inference about the tail. Second, tail observations are rare and sparse, making tail inference challenging even for large samples.

These difficulties are especially pronounced for non-ordinal discrete distributions. For a non-ordinal discrete probability distribution $(\Omega,\mathcal F,P)$, the sample space $\Omega$ is a set of possible outcomes without an inherent numerical scale. For example, $\Omega$ may be the vocabulary from which words are drawn, the set of webpages that may be visited, the set of species that may occur in a region, the set of product identifiers that may appear in transactions, or the set of nodes that may be contacted in a communication system. Although these outcomes are not ordered by magnitude, they can be ranked by their probabilities, yielding a ranked probability mass function. Tail inference in this setting concerns the behavior of the small probabilities associated with rare outcomes, rather than the behavior of large numerical observations. The qualitative class of ranked tail decay is therefore a first-order modeling question: geometric, or exponential-type, tails suggest that probabilities decrease rapidly with rank, whereas sub-exponential or power-law tails imply a substantially heavier contribution from rare categories.

Classical extreme-value methods provide powerful tools for tail inference once the relevant tail representation has been specified. Hill's estimator, for instance, gives a nonparametric estimate of Pareto-type, or regularly varying, tail behavior \cite{hill1975simple}. Subsequent work has refined this line of inference through studies of optimal rates for estimating parameters of regular variation \cite{hall1984best}, generalized and moment-type estimators of the extreme-value index \cite{dekkers1989moment,alves1995estimation}, threshold-exceedance methods based on generalized Pareto approximations \cite{smith1987estimating}, and semiparametric corrections for departures from exact Pareto behavior \cite{feuerverger1999estimating}. \textcolor{black}{Alternative approaches estimate power-law tail indices through scaling properties rather than directly through upper order statistics. For example, \cite{crovella1999estimating} proposed a nonparametric scaling estimator based on the aggregation behavior of sums of heavy-tailed random variables, together with graphical diagnostics for identifying the relevant scaling region. Like Hill-type procedures, such scaling-based methods are designed primarily for power-law tails and require the analyst to identify a range over which the corresponding asymptotic scaling behavior is informative.} A second practical issue is the choice of where the tail analysis should begin: tail estimators depend critically on the number of upper order statistics, or sample fraction, used in estimation. This problem is studied by Drees and Kaufmann \cite{drees1998selecting}, among others. A comprehensive treatment of the probabilistic and statistical foundations of extreme-value theory can be found in \cite{de2006extreme}.

The present work, the Turing--Entropic Tail Classifier (TENT), approaches tail inference from a different but related direction. Rather than starting from extreme order statistics, TENT is based on recent developments in domain-of-attraction theory on countable alphabets \cite{zhang2018domains}, a line of work connected to Turing's formula \cite{good1953population} and generalized Simpson-type indices \cite{simpson1949measurement, grabchak2017generalized}. Turing's formula relates the prevalence of rare symbols in a sample to the probability mass of unseen or low-frequency symbols, suggesting that frequency-of-frequency statistics can reveal information about the decay of ranked probabilities. Building on this viewpoint, TENT organizes information-theoretic functionals of empirical symbol frequencies into a tail profile and compares this profile with benchmarks corresponding to several ranked-tail regimes. Under a power-law assumption, the resulting inference is closely related to Hill-type tail-index estimation, revealing an intrinsic connection between the alphabet-based and classical extreme-value viewpoints. At the same time, the tail-profile approach is not restricted to the power-law case: in a manner reminiscent of extreme-value-index methods, it distinguishes thin-tailed and thick-tailed behavior, while further separating exponential-type, near-exponential, sub-exponential, and power-law regimes. Like classical tail-index procedures, TENT also faces the practical problem of determining which portion of the empirical tail should be used for reliable inference. This article provides practical guidance for this choice, while a systematic theory of optimal tail-region selection is left for future work.

The rest of the article is organized as follows. Section \ref{sec-main} introduces TENT and presents the main theoretical results. Section \ref{sec-amazon} illustrates the use of TENT through an example based on Amazon stock return data, demonstrating its application to continuous observations through discretization. Section \ref{sec-simu} provides numerical studies, including a reanalysis of Hill's example data and simulation studies evaluating the performance of TENT under various distributions. Section \ref{sec-how-to-select-tail-profile} discusses the selection of tail profiles in practical applications. Section \ref{sec-conclusion} concludes the article. All proofs are provided in Appendix \ref{sec-proofs}.

\section{Main Results}
\label{sec-main}

This section introduces the main components of the Turing--Entropic Tail Classifier (TENT). The central object is the tail profile, a collection of probability-weighted functionals designed to summarize the decay of the ranked probability mass function. Its construction is motivated by Turing's formula, whose connection to missing probability suggests that low-frequency empirical symbols carry information about the unseen, and hence tail, portion of the distribution. After defining the tail profile, Theorem \ref{th2} shows that, for several representative tail families, the divergence rates of the tail profile identify the qualitative tail class and, in certain cases, associated tail parameters. Theorem \ref{th1} then establishes an almost surely consistent estimator of the tail profile. These results motivate a sequence of diagnostic plots, referred to as entropic plots, which form the basis of the proposed classifier.

\subsection{Tail Profile}
\label{subsec-tail-profile}

Let $\mathbf{p}=\{p_k;k\geq 1\}$ be a probability distribution on a countably infinite alphabet $\mathscr{X}=\{\ell_k;k\geq 1\}$, where the elements of $\mathscr{X}$ are indexed in decreasing order of probabilities, so that $p_k=P(\{\ell_k\})$ and $p_k\geq p_{k+1}$ for all $k$. Let $\{X_i;i=1,\ldots,n\}$ be an independent and identically distributed $(iid)$ sample from $\mathbf{p}$. For each $k\geq 1$, let
\[
Y_k=\sum_{i=1}^n 1[X_i=\ell_k],
\qquad
\hat p_k=\frac{Y_k}{n},
\]
denote the empirical frequency and relative frequency of the symbol $\ell_k$. For $r=1,\ldots,n$, define the frequency-of-frequency counts
\[
N_r=\sum_{k\geq 1}1[Y_k=r],
\]
and let
\[
\pi_0=\sum_{k\geq 1}p_k 1[Y_k=0]
\]
denote the total probability mass of symbols not observed in the sample. The quantity $\pi_0$ is commonly called the missing probability, and $1-\pi_0$ is known as the sample coverage of the population, or simply the coverage; see, for example, \cite{chao1992estimating}.

The Good--Turing estimator of the missing probability is
\[
T=\frac{N_1}{n}.
\]
Often referred to as Turing's formula, this estimator was introduced by \cite{good1953population} and is widely attributed to Alan Turing. It estimates the total probability mass assigned to symbols that are not represented in an independent and identically distributed sample of size $n$. Its statistical properties have been studied extensively; see, among others, \cite{robbins1968estimating, esty1983normal, chao1988generalized, zhang2008sufficient, zhang2009asymptotic}. A comprehensive introduction to this line of work is provided in \cite{zhang2016statistical}.

Its connection to the proposed tail profile is revealed by its expectation \cite{robbins1968estimating}. Since
\[
\mathbb{E}(N_1)
=
\sum_{k\geq 1}\mathbb{P}(Y_k=1)
=
\sum_{k\geq 1} n p_k(1-p_k)^{n-1},
\]
it follows that
\[
\mathbb{E}(T)
=
\sum_{k\geq 1}p_k(1-p_k)^{n-1}.
\]
Thus, the Good--Turing statistic is centered around a probability-weighted functional that emphasizes symbols with probabilities of order $1/n$. This form connects low-frequency observed symbols with unseen probability mass and is closely related to generalized Simpson-type functionals. Motivated by this structure, the following definition introduces the tail index and tail profile.

\begin{definition}[Tail index and tail profile]
Let
\[
\zeta_v=\sum_{k\geq 1}p_k(1-p_k)^v.
\]
The quantity
\[
\tau_v=v\zeta_v
\label{tauv}
\]
is referred to as the \textbf{tail index}. For positive integers $v_1<v_2$, the collection
\[
\{\tau_v; v=v_1,\ldots,v_2\}
\]
is referred to as a \textbf{tail profile}.
\label{def-tail}
\end{definition}

The quantity $\zeta_v$ is referred to as the \textbf{entropic basis}. This terminology comes from its connection to Shannon's entropy \cite{zhang2012entropy}. $\zeta_v$ forms a natural basis for decomposing entropy into probability-weighted components. Moreover, $\zeta_v$ is a special case of generalized Simpson-type indices \cite{grabchak2017generalized}.

The parameter $v$ determines which part of the ranked mass function is emphasized. Smaller values of $v$ may still reflect the body of the distribution, whereas larger values place more weight on the small-probability region through the factor $(1-p_k)^v$. Thus, the choice of $v$ involves a practical trade-off: choosing $v$ too small may obscure the tail behavior, while choosing $v$ too large may lead to unstable estimation because the relevant symbols are rarely observed. This challenge is analogous to the sample-fraction or threshold-selection problem in Hill-type tail-index estimation, as studied, for example, by Drees and Kaufmann \cite{drees1998selecting}. The scaled quantity $\tau_v=v\zeta_v$ is used because it has tractable asymptotic behavior for the benchmark tail families studied below.


\subsection{Properties of the Tail Profile}
\label{subsec-tail-profile-properties}

The usefulness of the tail profile comes from the fact that the asymptotic behavior of $\tau_v$ reflects the decay rate of the ranked probability mass function. This connection has its roots in domain-of-attraction theory on countable alphabets. In particular, \cite{zhang2018domains} studied the behavior of $\tau_v$ for different classes of distributions on countable alphabets and showed that distinct tail regimes lead to distinct asymptotic patterns of $\tau_v$.

A first distinction is between thick-tailed and thin-tailed ranked distributions. For slowly decaying ranked probabilities, $\tau_v$ tends to diverge as $v\to\infty$. The following sufficient condition, from Theorem 6.3 of \cite{zhang2016statistical}, gives a simple way to identify such behavior.

\begin{prop}[Theorem 6.3 in \cite{zhang2016statistical}]
If $p_{k+1}/p_k\to 1$, then $\tau_v\to\infty$ as $v\to\infty$.
\label{prop-pratio}
\end{prop}

The condition above is satisfied by several standard ranked tail families, including the following:
\begin{align}
\text{Power tail:} \quad 
p_k &\propto k^{-\lambda}, 
&& k\geq k_0,\quad \lambda>1, \quad k_0\geq 1; 
\label{dist1}\\
\text{Sub-exponential tail:} \quad 
p_k &\propto \exp(-\lambda k^\alpha), 
&& k\geq k_0,\quad \lambda>0,\quad \alpha\in(0,1),\quad k_0\geq 1;
\label{dist2}\\
\text{Near-exponential tail:} \quad 
p_k &\propto \exp\left(-\frac{\lambda k}{\ln^\beta k}\right),
&& k\geq k_0,\quad \lambda>0,\quad \beta>0,\quad k_0\geq 1.
\label{dist3}
\end{align}
Here $a_k\propto b_k$ means $a_k=c b_k$ for a positive constant $c$.

By contrast, if $p_{k+1}/p_k\to c<1$, the ranked probabilities decay at an exponential-type rate. Examples include
\[
p_k\propto e^{-\lambda k}
\qquad\text{and}\qquad
p_k\propto e^{-\lambda k^2},
\]
for $k\geq k_0$ and $\lambda>0$. Such distributions belong to the Molchanov family in the terminology of \cite{zhang2018domains}, for which $\tau_v$ remains bounded and oscillates between positive constants rather than converging to infinity. Thus, divergence of $\tau_v$ provides a useful separation between thick-tailed ranked distributions and exponential-type or thinner ranked distributions.

The following theorem shows that the three representative thick-tailed families in \eqref{dist1}--\eqref{dist3} have distinguishable asymptotic signatures in terms of $\tau_v$. These signatures provide the basis for their use as benchmark regimes in the proposed classifier.

\begin{theorem}
For a distribution $\mathbf p$ in the form of \eqref{dist1}, \eqref{dist2}, or \eqref{dist3}, the corresponding tail index $\tau_v$ diverges as $v\to\infty$ at the following rates:
\label{th2}
\begin{align}
\text{Power tail:} 
\quad & \tau_v\asymp v^{1/\lambda}, 
&&\lambda>1; \label{powerpi}\\[3pt]
\text{Sub-exponential tail:} 
\quad & \tau_v\asymp (\ln v)^{1/\alpha-1}, 
&&\alpha\in(0,1); \label{subexppi}\\[3pt]
\text{Near-exponential tail:} 
\quad & \tau_v\asymp (\ln\ln v)^\beta, 
&&\beta>0. \label{nearexppi}
\end{align}
Here $f(v)\asymp g(v)$ means that there exist constants $0<c_1<c_2<\infty$ such that
\[
c_1g(v)\leq f(v)\leq c_2g(v)
\]
for all sufficiently large $v$.
\end{theorem}

The proof is provided in Appendix \ref{subsec-proof-main}. The three rates in Theorem \ref{th2} suggest different linearizing transformations. Under a power tail, $\ln \tau_v$ is asymptotically linear in $\ln v$; under a sub-exponential tail, it is asymptotically linear in $\ln\ln v$; and under a near-exponential tail, it is asymptotically linear in $\ln\ln\ln v$. These distinguishable population-level signatures motivate the classifier. The next subsection turns to the estimation of $\tau_v$ from sample data.

\subsection{Estimation of the Tail Profile}
\label{subsec-tail-profile-estimation}

Theorem \ref{th2} concerns the population tail profile. To use these population-level signatures with data, one needs an empirical counterpart of $\tau_v$. Such an estimator is available through a $U$-statistic construction \cite{zhang2010re}.

For every positive integer $v\leq n-1$, define
\[
Z_v
=
\frac{n^{v+1}[n-(v+1)]!}{n!}
\sum_{k=1}^{\infty}
\left[
\hat p_k
\prod_{j=0}^{v-1}
\left(1-\hat p_k-\frac{j}{n}\right)
\right].
\]
Proposed by \cite{zhang2010re}, $Z_v$ is a $U$-statistic constructed from Turing's formula as a kernel of degree $v$. It satisfies
\[
\mathbb E(Z_v)=\zeta_v.
\]
Letting
\[
T_v=vZ_v,
\]
it follows that
\[
\mathbb E(T_v)=\tau_v.
\]
The empirical tail profile is therefore
\[
\{T_v; v\},
\]
which estimates the population tail profile
\[
\{\tau_v; v\}.
\]

The following theorem establishes that this empirical tail profile is consistent for each fixed $v$. The proof is provided in Appendix \ref{subsec-proof-main}.

\begin{theorem}
Let $\tau_v$ and $T_v$ be defined as above, with $v\leq n-1$. Then, for every fixed $v$,
\[
\frac{T_v}{\tau_v}\stackrel{a.s.}{\longrightarrow}1 \textcolor{black}{,}
\qquad\text{as } n\to\infty.
\]
\label{th1}
\end{theorem}

Theorem \ref{th1} justifies replacing the population tail profile $\{\tau_v\}$ by the empirical tail profile $\{T_v\}$ when examining the asymptotic signatures described in Theorem \ref{th2}. The next subsection describes how these signatures are used in practice through a sequence of entropic plots.

\subsection{Entropic Plots}
\label{subsec-classifier-plots}

Theorems \ref{th2} and \ref{th1} together motivate a graphical implementation of the classifier. Theorem \ref{th2} gives distinguishable population-level signatures for the benchmark tail regimes, while Theorem \ref{th1} justifies estimating the population tail profile $\{\tau_v\}$ by the empirical tail profile $\{T_v\}$. The resulting diagnostic displays are referred to as \textbf{entropic plots}, because they are based on the entropic basis $\{\zeta_v\}$; see also \cite{zhang2022entropic}.

\begin{figure}[t]
    \centering
    \includegraphics[width=0.6\linewidth]{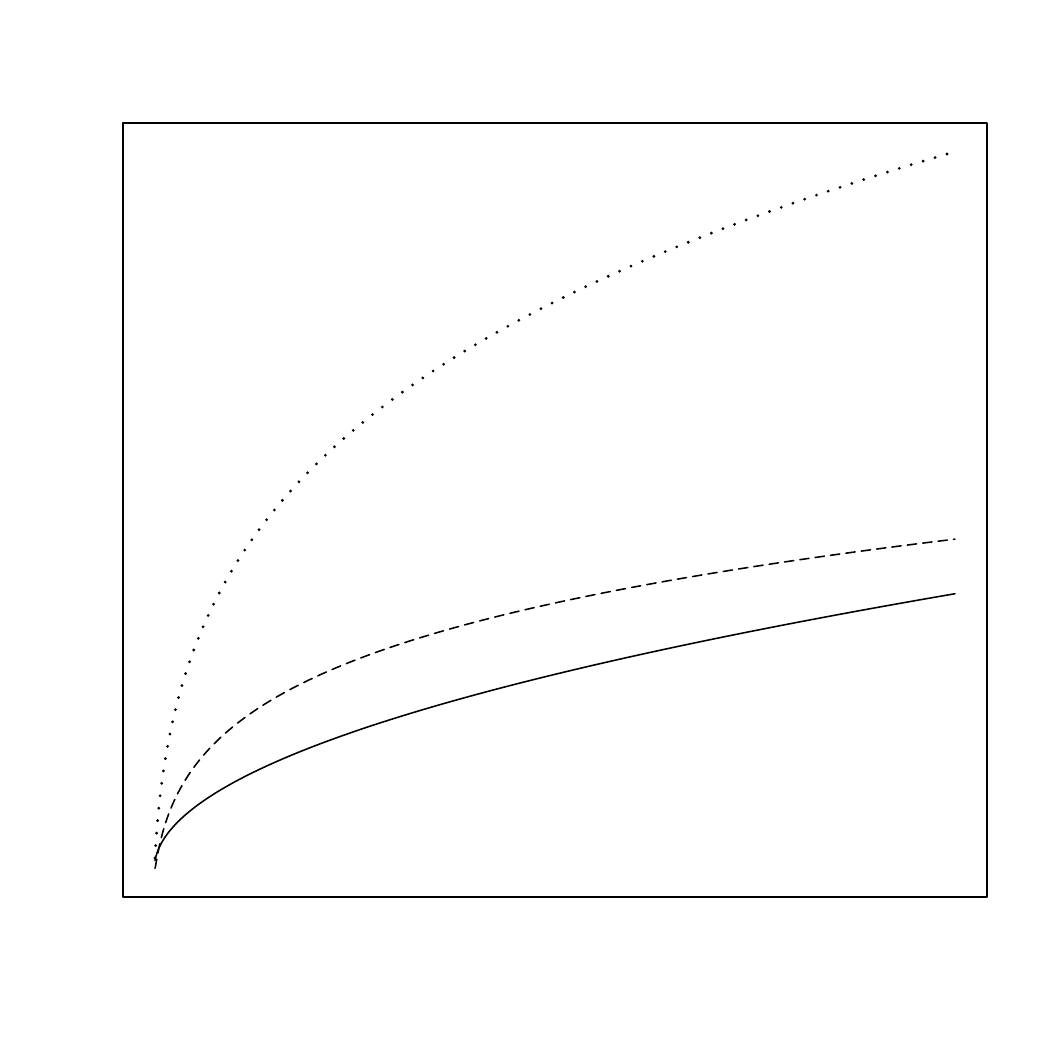}
    \caption{Population entropic Plot~0: $\tau_v$ versus $v$. The solid curve corresponds to power decay, the dashed curve to sub-exponential decay, and the long-dashed curve to near-exponential decay.}
    \label{fig1a}
\end{figure}

At the population level, the first plot, referred to as Plot~0, displays $\tau_v$ versus $v$. For the three representative thick-tailed families in \eqref{dist1}--\eqref{dist3}, Theorem \ref{th2} implies that $\tau_v\to\infty$ as $v\to\infty$. Moreover, the corresponding growth is sublinear in the sense reflected by the rates in \eqref{powerpi}--\eqref{nearexppi}, producing a characteristic increasing and concave pattern for sufficiently large $v$. Figure \ref{fig1a} gives qualitative sketches of this behavior for the three benchmark regimes.

Plot~0 is also useful for separating thick-tailed behavior from exponential-type or thinner tails. As discussed above, for distributions in the Molchanov family, including exponential and thinner ranked tails, $\tau_v$ does not diverge to infinity but instead remains bounded and oscillates between positive constants. Thus, when $v$ and the plotting range $(v_1,v_2)$ are sufficiently large, the profile may exhibit a downturn or oscillatory behavior. Such a pattern provides evidence against the thick-tailed regimes in \eqref{dist1}--\eqref{dist3} and suggests an exponential-type or thinner tail.

\begin{figure}[t]
    \centering
    \includegraphics[width=0.6\linewidth]{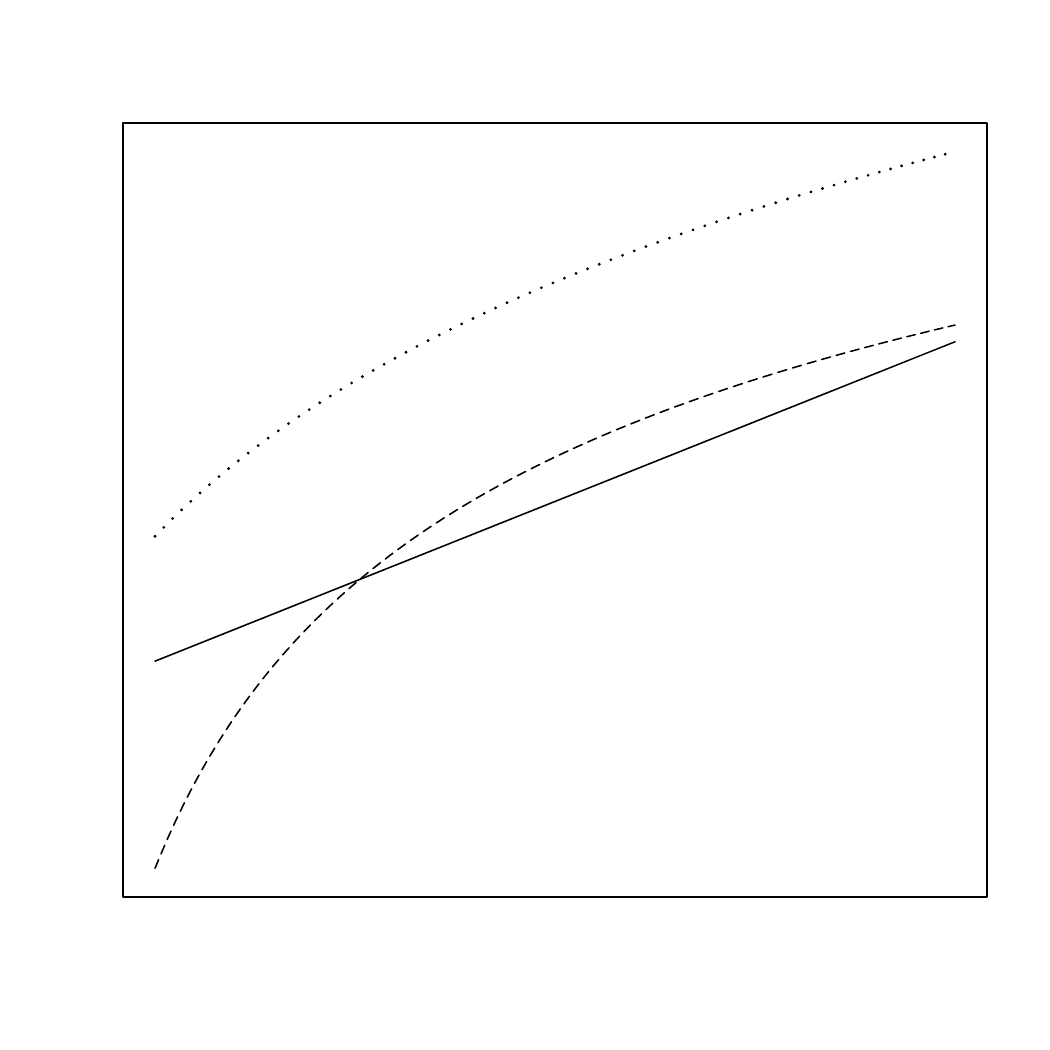}
    \caption{Population entropic Plot~1: $\ln(\tau_v)$ versus $\ln(v)$. A power tail produces the strongest linear pattern in this transformation.}
    \label{fig1b}
\end{figure}

\begin{figure}[t]
    \centering
    \includegraphics[width=0.6\linewidth]{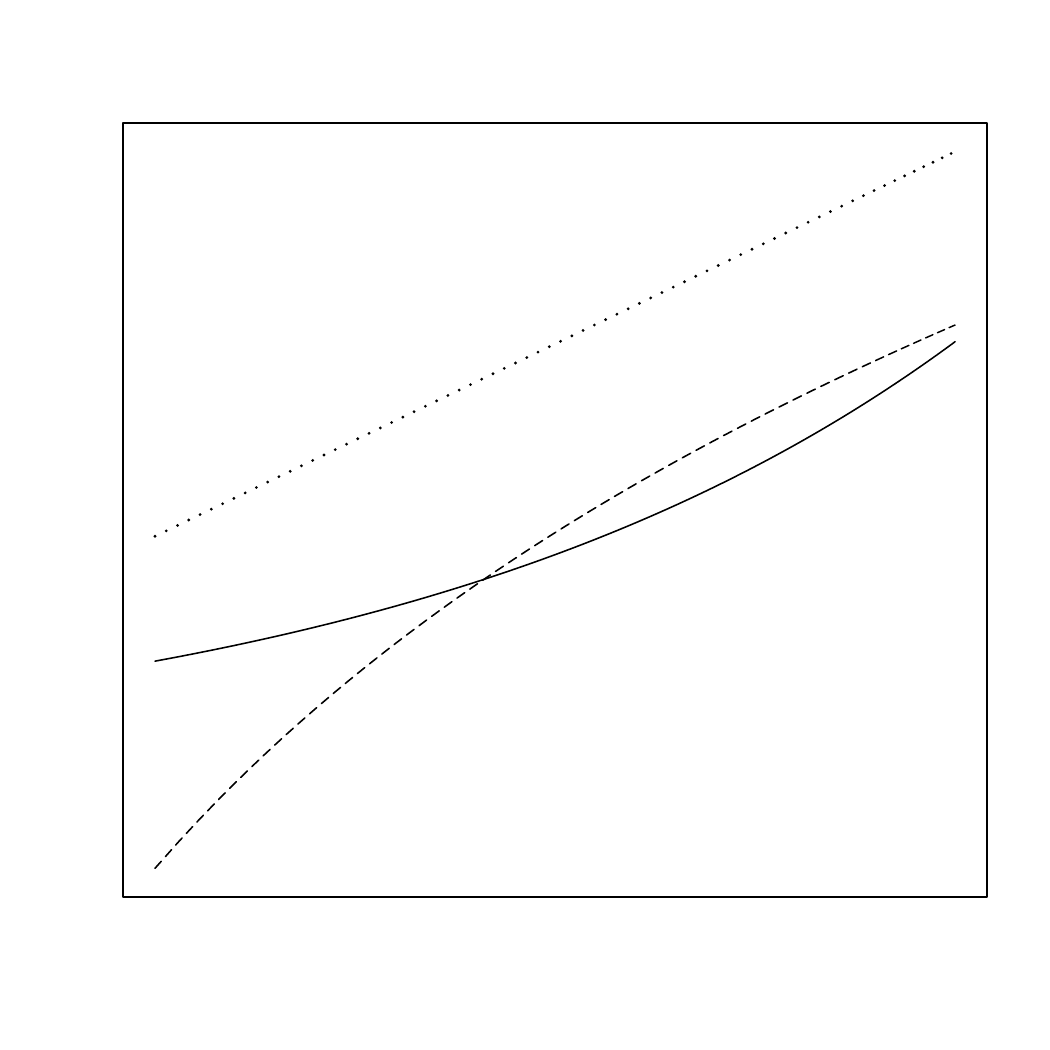}
    \caption{Population entropic Plot~2: $\ln(\tau_v)$ versus $\ln\ln(v)$. A sub-exponential tail produces the strongest linear pattern in this transformation.}
    \label{fig1c}
\end{figure}

\begin{figure}[t]
    \centering
    \includegraphics[width=0.6\linewidth]{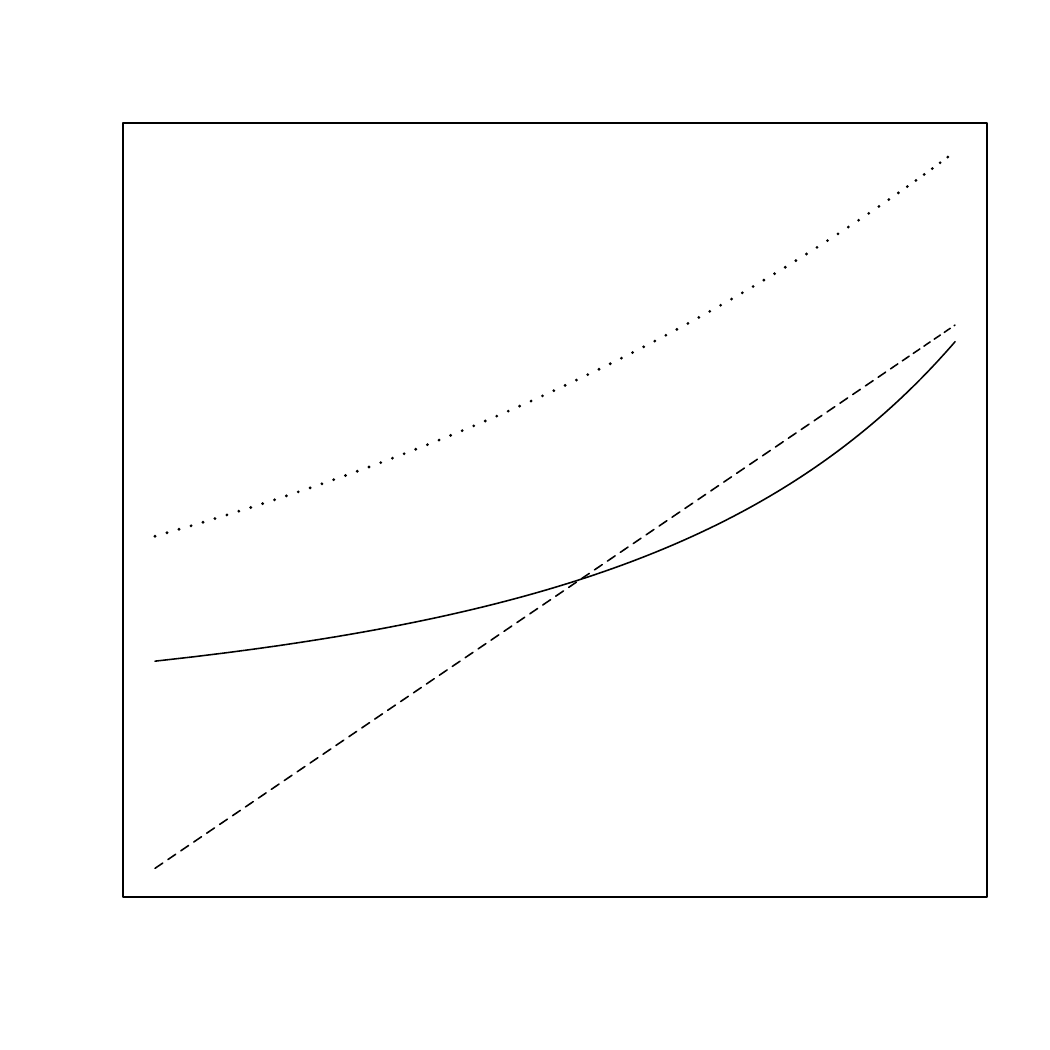}
    \caption{Population entropic Plot~3: $\ln(\tau_v)$ versus $\ln\ln\ln(v)$. A near-exponential tail produces the strongest linear pattern in this transformation.}
    \label{fig1d}
\end{figure}

The next three plots are designed to distinguish among the three thick-tailed benchmark regimes. Plot~1 displays $\ln \tau_v$ versus $\ln v$; Plot~2 displays $\ln \tau_v$ versus $\ln\ln v$; and Plot~3 displays $\ln \tau_v$ versus $\ln\ln\ln v$. By Theorem \ref{th2}, these transformations linearize different tail regimes:
\[
\begin{array}{lll}
\text{Power tail:} & \ln \tau_v \text{ is asymptotically linear in } \ln v, \\
\text{Sub-exponential tail:} & \ln \tau_v \text{ is asymptotically linear in } \ln\ln v, \\
\text{Near-exponential tail:} & \ln \tau_v \text{ is asymptotically linear in } \ln\ln\ln v.
\end{array}
\]
Accordingly, among Plots~1--3, the plot exhibiting the clearest linear trend provides evidence for the corresponding tail regime. These qualitative features are illustrated in Figures \ref{fig1b}--\ref{fig1d}.

In practice, the population profile $\{\tau_v\}$ is unknown and is replaced by the empirical profile $\{T_v\}$. The empirical entropic plots therefore display
\[
T_v \text{ versus } v,\qquad
\ln T_v \text{ versus } \ln v,\qquad
\ln T_v \text{ versus } \ln\ln v,\qquad
\ln T_v \text{ versus } \ln\ln\ln v,
\]
respectively, over a selected range $v_1\leq v\leq v_2$. The selection of this range affects the stability and interpretability of the plots, and practical guidance for choosing the tail profile is discussed in Section \ref{sec-how-to-select-tail-profile}. Section \ref{sec-amazon} illustrates the use of the classifier with real data.

\section{A Real Data Example: Amazon Stock Returns}
\label{sec-amazon}

This section illustrates the use of TENT with Amazon (AMZN) minute log-return data. Although TENT is developed for discrete distributions, continuous observations can be brought into the framework through discretization; see Theorem \ref{thm-bin_tail_type_same} in Appendix \ref{subsec-proof-bin}. The example is intended to demonstrate the workflow of the classifier by separately analyzing the discretized left and right tails of the log-return distribution. The classifier is implemented in R using the \texttt{TailClassifier} package (version 0.1.2 in R 4.5.2; see \cite{r-TailClassifier}).

The data consist of 30,000 minute log returns collected from 09:31 on 03/13/2007 to 15:56 on 06/29/2007. The left-tail data are obtained by selecting the negative log returns and discretizing their magnitudes using a common bin width of \(10^{-4}\). The choice of bin width is not unique; however, Theorem \ref{thm-bin_tail_type_same} suggests that the tail type is preserved under common-width discretization, under suitable conditions. As a sensitivity check, Appendix \ref{subsec-proof-bin} reports results using a coarser bin width of \(10^{-3}\), which are comparable to those obtained with bin width \(10^{-4}\); see Tables \ref{r-box-l3} and \ref{r-box-r3}. The right-tail data are obtained similarly from the positive log returns. After discretization, the left-tail sample contains \(n=12,523\) observations, and the right-tail sample contains \(n=12,968\) observations.

To apply the classifier, one first chooses a range of indices \(v_1\leq v\leq v_2\) over which the empirical tail profile is examined. Preliminary entropic plots are used to identify a stable and interpretable range. \textcolor{black}{The selection of this range is discussed further in Section~\ref{sec-how-to-select-tail-profile}.}

\begin{figure}[htbp]
    \centering
    \includegraphics[width=\linewidth]{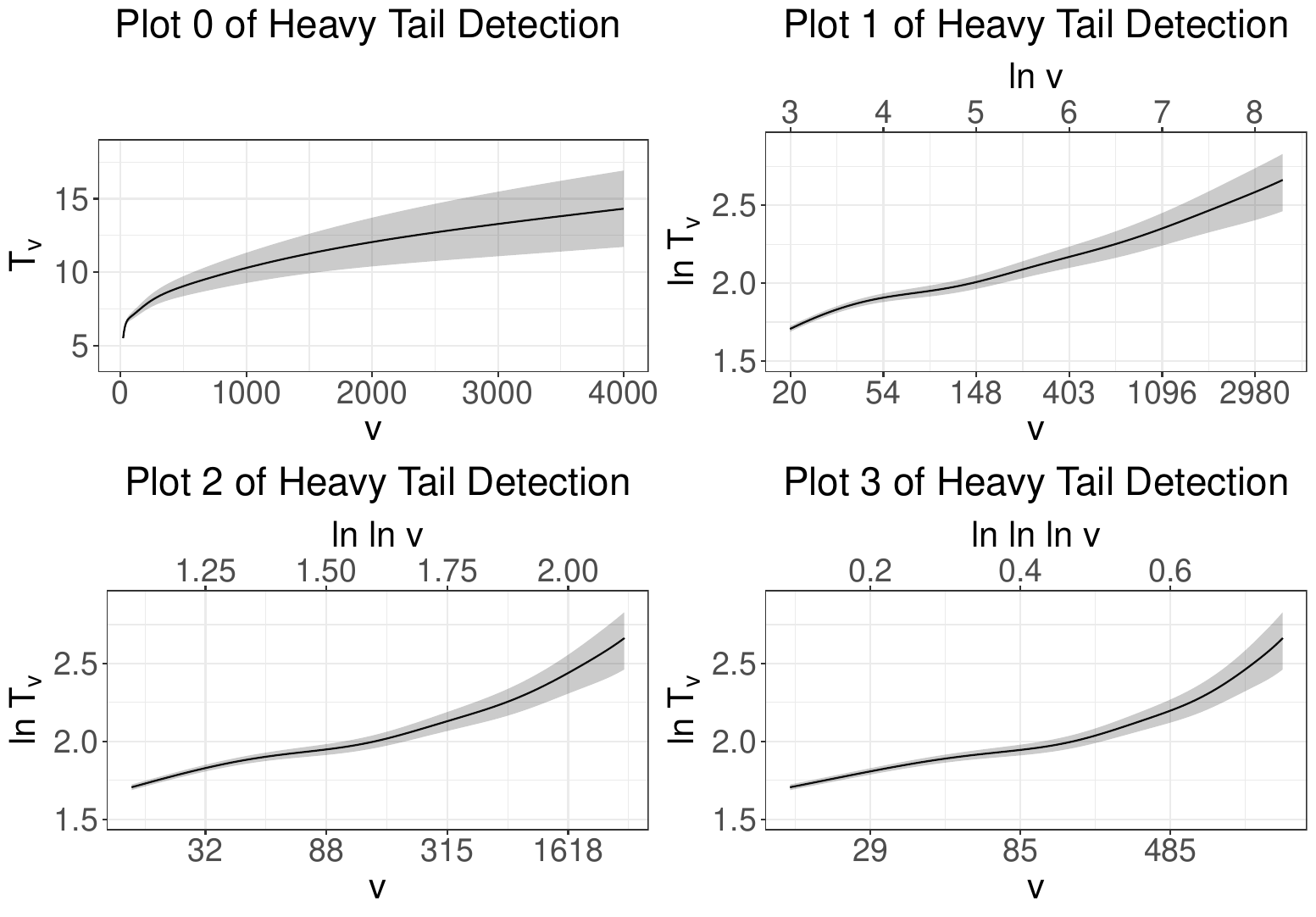}
    \caption{Preliminary entropic plots for AMZN left-tail minute log-return data with bin width \(10^{-4}\).}
    \label{fig-amzn_l4}
\end{figure}

Figure~\ref{fig-amzn_l4} shows the preliminary entropic plots for the left-tail data. The confidence bands are derived from the asymptotic normal approximation in Lemma~\ref{lemma-Tv-normal}; see Appendix~\ref{subsec-proof-CI}. In Plot~0, the upward trend suggests a tail thicker than exponential. In Plots~1, 2, and 3, apparent inflection points occur near the integer indices \(v=90\), \(v=111\), and \(v=121\), corresponding respectively to \(\floor{e^{4.5}}\), \(\floor{\exp(e^{1.55})}\), and \(\floor{\exp(\exp(e^{0.45}))}\). Since such inflection points are not expected in the large-\(v\) population signatures described in Theorem~\ref{th2}, the portions beyond these values are treated as unstable for inference.

\begin{figure}[htbp]
    \centering
    \includegraphics[width=\linewidth]{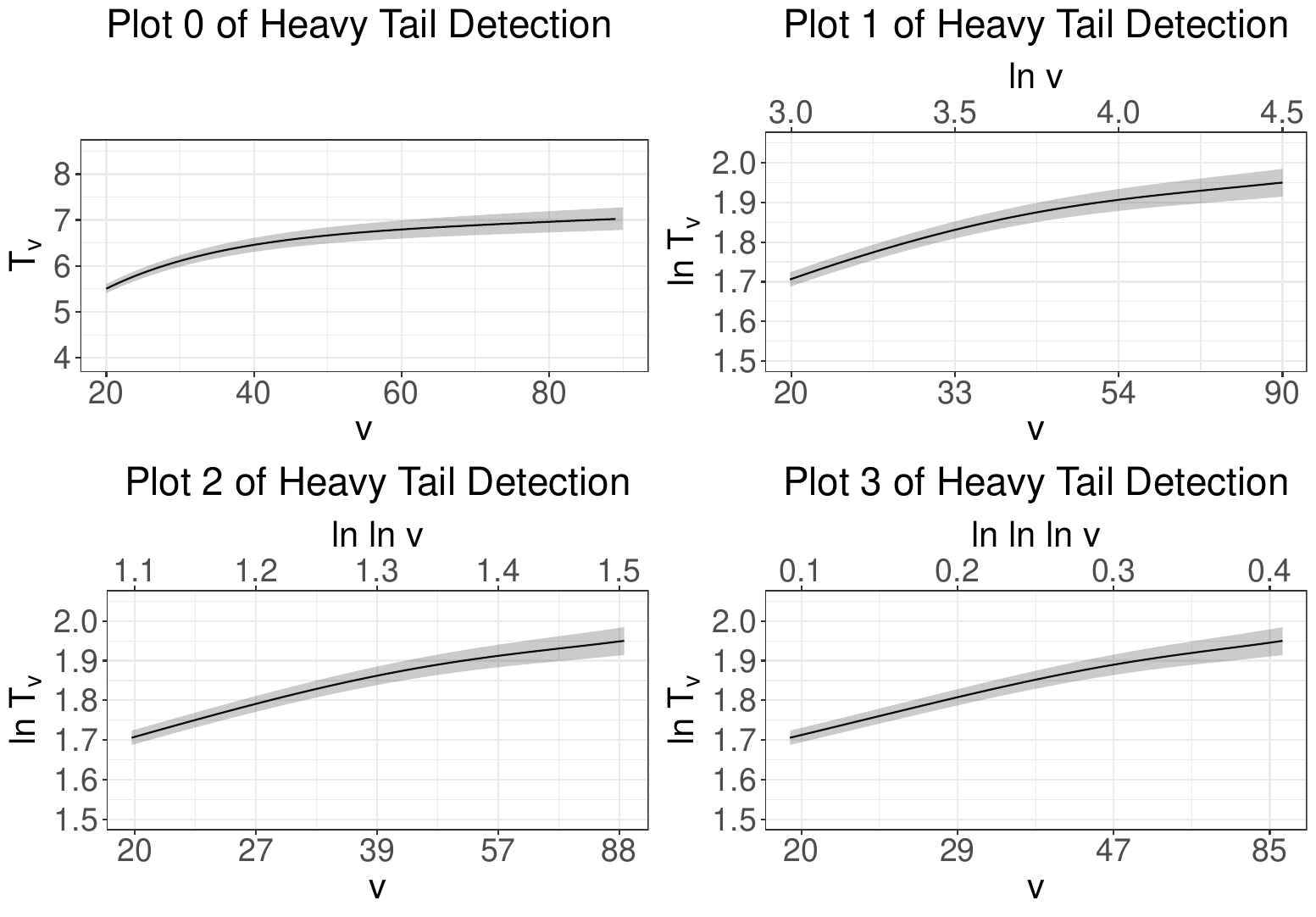}
    \caption{Refined entropic plots with \(v\leq 90\) for AMZN left-tail minute log-return data with bin width \(10^{-4}\).}
    \label{fig-amzn_l4_2}
\end{figure}

Figure~\ref{fig-amzn_l4_2} displays the refined entropic plots with \(v\leq 90\). The upward trend in Plot~0 continues to suggest a tail thicker than exponential. Among Plots~1, 2, and 3, Plot~3 exhibits the clearest linear trend, suggesting a near-exponential tail. For parameter estimation, the tail profile from
\[
v=\floor{\exp(\exp(e^{0.2}))}=29
\qquad\text{to}\qquad
v=\floor{\exp(\exp(e^{0.3}))}=47 
\]
is selected; practical considerations for tail-profile selection are discussed in Section~\ref{sec-how-to-select-tail-profile}. The corresponding output is shown in Table~\ref{r-box-l4}. The near-exponential classification is based on the highest Pearson correlation coefficient among the three linearized plots. The fitted form is
\[
p_k\propto \exp\left\{-\frac{\lambda k}{(\ln k)^{0.83}}\right\},
\]
with \(\hat{\beta}=0.83\) and a 95\% confidence interval for \(\beta\) of \((0.39,1.27)\). Details on the confidence interval are provided in Appendix~\ref{subsec-proof-CI}.

\begin{rcodeboxtable}{R output of \texttt{TailClassifier()} for left-tail data with bin width \(\delta=10^{-4}\).}{r-box-l4}
\begin{lstlisting}
$Conclusion
[1] "The data suggest a near-exponential decaying tail."
$CI
[1] "95% CI based on information from v_left = 29 to v_right = 47"
    Tail_Type          CI_level    lwr         upr         PointEstimate
1   Power              0.95        5.2488340   11.8223654  5.5642004
2   Sub-Exponential    0.95        0.5028738   0.7664072   0.6069464
3   Near-Exponential   0.95        0.3901127   1.2653127   0.8288801
\end{lstlisting}
\end{rcodeboxtable}

The same procedure is applied to the right-tail data.

\begin{figure}[htbp]
    \centering
    \includegraphics[width=\linewidth]{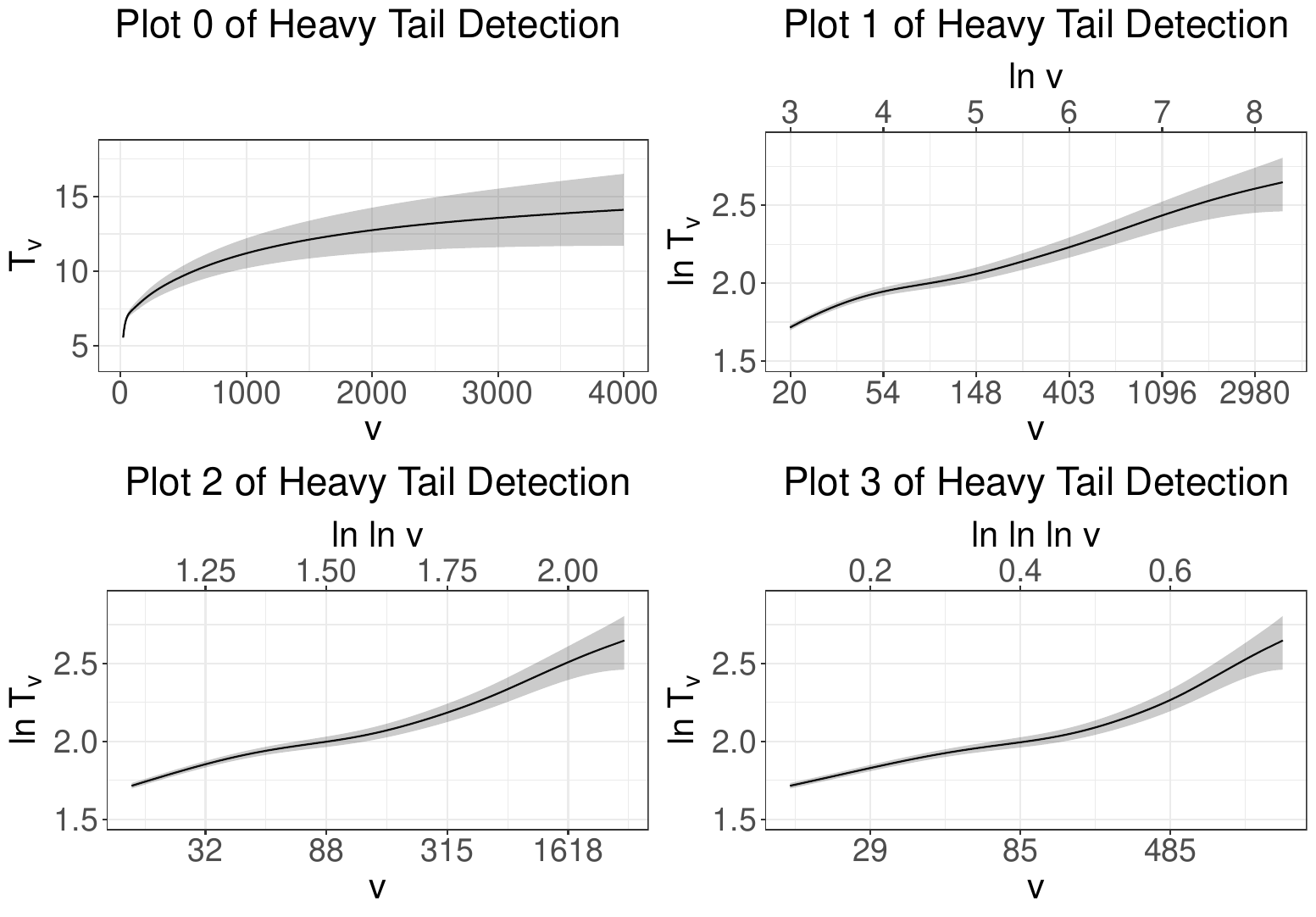}
    \caption{Preliminary entropic plots for AMZN right-tail minute log-return data with bin width \(10^{-4}\).}
    \label{fig-amzn_r4}
\end{figure}

Figure~\ref{fig-amzn_r4} shows the preliminary entropic plots for the right-tail data. In Plots~1, 2, and 3, apparent inflection points occur near the integer indices \(v=90\), \(v=111\), and \(v=121\), corresponding respectively to \(\floor{e^{4.5}}\), \(\floor{\exp(e^{1.55})}\), and \(\floor{\exp(\exp(e^{0.45}))}\). To remove the unstable portions of the profile, the refined plots are restricted to \(v\leq 90\).

\begin{figure}[htbp]
    \centering
    \includegraphics[width=\linewidth]{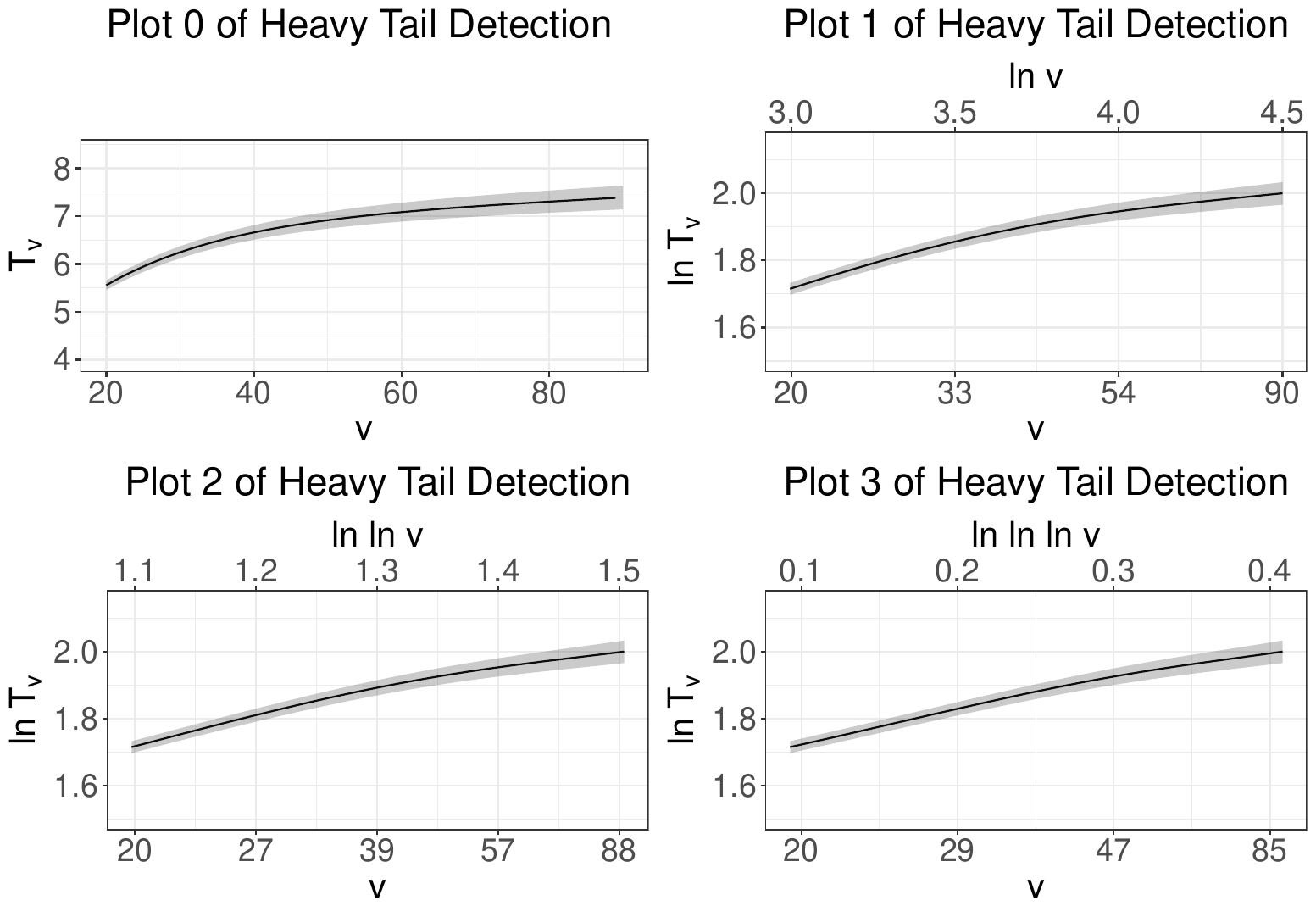}
    \caption{Refined entropic plots with \(v\leq 90\) for AMZN right-tail minute log-return data with bin width \(10^{-4}\).}
    \label{fig-amzn_r4_2}
\end{figure}

Figure~\ref{fig-amzn_r4_2} displays the refined entropic plots for the right-tail data. The upward trend in Plot~0 suggests a tail thicker than exponential, and Plot~3 exhibits the clearest linear trend among Plots~1--3. The right tail is therefore classified as near-exponential. Using the same tail profile from \(v=29\) to \(v=47\), the output is shown in Table~\ref{r-box-r4}. The fitted form is
\[
p_k\propto \exp\left\{-\frac{\lambda k}{(\ln k)^{0.97}}\right\},
\]
with \(\hat{\beta}=0.97\) and a 95\% confidence interval for \(\beta\) of \((0.54,1.39)\).

\begin{rcodeboxtable}{R output of \texttt{TailClassifier()} for right-tail data with bin width \(\delta=10^{-4}\).}{r-box-r4}
\begin{lstlisting}
$Conclusion
[1] "The data suggest a near-exponential decaying tail."
$CI
[1] "95% CI based on information from v_left = 29 to v_right = 47"
    Tail_Type          CI_level    lwr         upr         PointEstimate
1   Power              0.95        4.5543854   8.5253736   4.7700106
2   Sub-Exponential    0.95        0.4792621   0.7029088   0.5696663
3   Near-Exponential   0.95        0.5409798   1.3907094   0.9668856
\end{lstlisting}
\end{rcodeboxtable}

Regarding the horizontal axes in Plots~1, 2, and 3, the spacing may initially appear disproportionate if read in terms of the lower-axis labels. The actual plotting scales are the transformed indices: \(\ln v\) for Plot~1, \(\ln\ln v\) for Plot~2, and \(\ln\ln\ln v\) for Plot~3. These transformed scales are displayed on the upper horizontal axes. The lower horizontal axes report the corresponding original values of \(v\) only as a reference, so that the reader can see which portion of the tail profile is being used. Since the plots are not drawn on the original \(v\)-scale, the spacing of the lower-axis \(v\) labels is not expected to be uniform. The next section presents numerical studies that further demonstrate the use of the proposed method and evaluate its performance under controlled distributional settings.

\section{Numerical Studies}
\label{sec-simu}

The numerical studies are divided into two parts. Section~\ref{subsec-uscities} revisits the city-population example from \cite{hill1975simple} to compare TENT with classical Hill-type inference. Section~\ref{subsec-simu} presents simulation studies to assess the performance of TENT under controlled distributional settings.

\subsection{U.S. City Populations in 1950}
\label{subsec-uscities}

Hill \cite{hill1975simple} considered population data for U.S. cities from the \emph{Statistical Abstract of the United States} (1950, p.~57) as an illustration of Pareto-type tail inference. Under a power-law tail assumption, Hill reported tail-parameter estimates of \(\hat{\alpha}_0(r)=1.35\) and \(\hat{\alpha}_1(r)=1.40\). The purpose of the present example is to compare the behavior of TENT with Hill's analysis on a discretized version of the same data, while emphasizing that TENT does not impose the power-law assumption at the outset.

For this comparison, the city-population data are discretized into 10 equal-width bins, yielding observed bin frequencies
\[
\{189,4,2,0,1,0,0,0,0,1\}.
\]
The corresponding TENT output is reported in Table~\ref{table-UScity}. Because the sample is sparse after discretization, the resulting confidence intervals should be interpreted with caution. \textcolor{black}{The point estimate is obtained from the fitted slope using all values in the selected tail profile, whereas the confidence interval is constructed by transforming endpoint bounds derived from the confidence bands at \(v_1\) and \(v_2\). The interval is therefore not formed by adding and subtracting a margin of error from the point estimate and need not be centered at that estimate. Because the point estimate and interval are based on different finite-sample calculations, the point estimate may occasionally fall outside the reported interval, particularly for small or sparse samples. Such an occurrence indicates that the selected profile contains limited information for stable parameter estimation and should be treated as a diagnostic warning.}

\begin{rcodeboxtable}{R output of \texttt{TailClassifier()} for U.S. city populations in 1950.}{table-UScity}
\begin{lstlisting}
$Conclusion
[1] "The data suggest a sub-exponential decaying tail."
$CI
[1] "95% CI based on information from v_left = 16 to v_right = 25"
    Tail_Type          CI_level    lwr         upr         PointEstimate
1   Power              0.95        1.32812720  Inf         1.2197631
2   Sub-Exponential    0.95        0.08187629  1           0.2897336
3   Near-Exponential   0.95        0           12.25314    2.6787161
\end{lstlisting}
\end{rcodeboxtable}

Table~\ref{table-UScity} classifies the discretized data as sub-exponential when no power-law assumption is imposed in advance. This conclusion should not be interpreted as contradicting Hill's analysis. Rather, it reflects the different inferential questions addressed by the two methods: Hill's estimator estimates a tail parameter under a Pareto-type assumption, whereas TENT first attempts to classify the broad tail regime. Indeed, if attention is restricted to the power-tail component of the TENT output, Hill's estimates, \(1.35\) and \(1.40\), fall within the reported 95\% confidence interval \((1.328,\infty)\). Thus, conditional on a power-law interpretation, the two analyses are compatible.

At the same time, the confidence intervals in Table~\ref{table-UScity} reveal substantial uncertainty. The power-tail interval is unbounded above, the sub-exponential interval reaches the boundary value \(1\), and the near-exponential interval has lower bound \(0\). These features indicate that the discretized sample contains limited information for reliably distinguishing among the candidate tail regimes. This example therefore illustrates both the connection between TENT and Hill-type inference under a power-law assumption, and the additional role of TENT as a preliminary classifier when the appropriate tail regime is not assumed in advance. Further discussion of tail-profile selection and reliability diagnostics is provided in Section~\ref{sec-how-to-select-tail-profile}.

\subsection{Simulation Study}
\label{subsec-simu}

This subsection evaluates the finite-sample performance of TENT under three representative tail regimes: power, sub-exponential, and near-exponential decay. The simulation designs correspond to the three benchmark families in \eqref{dist1}--\eqref{dist3}. Specifically, the target tail parameters are set to \(1.5\) for the power distribution, \(0.5\) for the sub-exponential distribution, and \(2\) for the near-exponential distribution.

\textcolor{black}{The subsection is organized in two parts. The first part presents three illustrative examples, each based on a single simulated data set, to demonstrate the entropic plots, tail-profile selection, classification, and parameter estimation produced by TENT. These examples are intended to illustrate the application of the method rather than to provide a systematic simulation assessment. The second part reports a Monte Carlo study based on \(1,000\) additional samples from each distribution to evaluate classification accuracy, interval coverage, interval width, and parameter estimation.}

The three simulation settings are as follows:
\begin{itemize}
    \item[1.] \textbf{Power, or Zipf-type, distribution:}
    \[
    p_k \propto k^{-1.5}, \qquad k=1,2,3,\ldots.
    \]
    \item[2.] \textbf{Sub-exponential distribution:}
    \[
    p_k \propto \exp(-\sqrt{k}), \qquad k=1,2,3,\ldots.
    \]
    \item[3.] \textbf{Near-exponential distribution:}
    \[
    p_k \propto \exp\left(-\frac{k}{\ln^2 k}\right), \qquad k=2,3,\ldots.
    \]
\end{itemize}

The first two probability sequences are monotone decreasing. The near-exponential sequence is not necessarily monotone over its full support, but its ranked probabilities exhibit the intended near-exponential decay in the distant tail. This is the relevant feature for the ranked-tail analysis considered \textcolor{black}{in this article}.

\subsubsection*{\textcolor{black}{Illustrative Example: Power Distribution}}

The power-tail simulation is generated from the continuous density
\[
f(x)=\frac{1}{2}x^{-1.5}, \qquad x>1.
\]
A continuous random sample drawn from this density is converted into a discrete sample by taking the floor of each observation. By Theorem~\ref{thm-bin_tail_type_same}, the resulting discrete distribution preserves the power-law tail type, with tail parameter \(1.5\).

\begin{figure}[htbp]
    \centering
    \includegraphics[width=\linewidth]{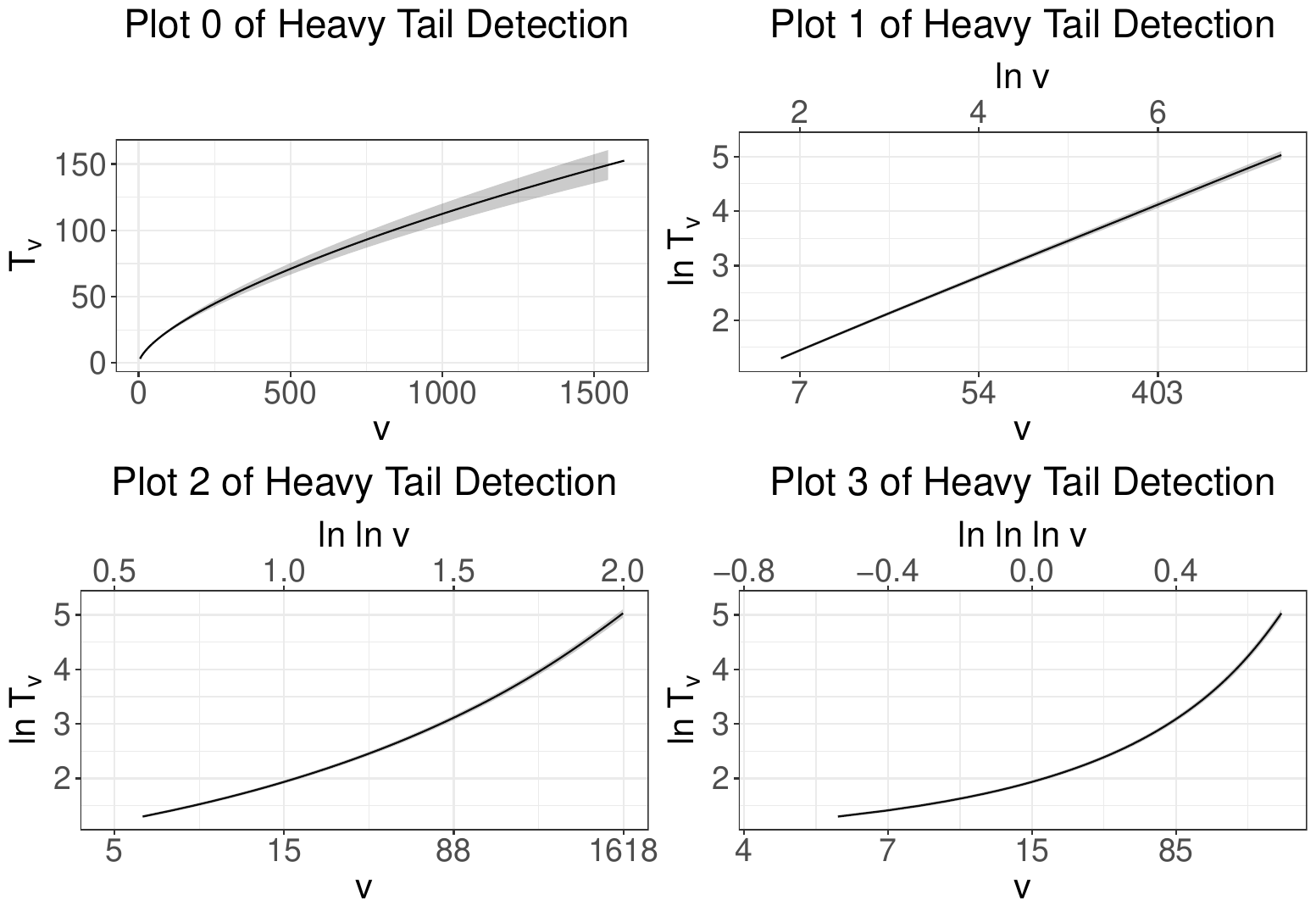}
    \caption{Entropic plots for a \textcolor{black}{single} random sample of size \(4,000\) generated from a power-decaying tail distribution with parameter \(1.5\).}
    \label{fig-pareto0.5}
\end{figure}

A discrete random sample of size \(4,000\) was generated using this \textcolor{black}{data-generating model} with random seed \(2023\). The corresponding entropic plots, along with 95\% confidence bands, are shown in Figure~\ref{fig-pareto0.5}. In Plot~0, the upward trend indicates a tail thicker than exponential. Among Plots~1, 2, and 3, Plot~1 exhibits the clearest linear trend, suggesting power-law decay. A tail profile from \(v=\floor{e^3} = 20\) to \(v= \floor{e^4} = 54\) was selected for parameter estimation. The output is reported in Table~\ref{r-box-Pareto0.5}, where the 95\% confidence interval \((1.4815,1.6779)\) contains the true value \(1.5\).

\begin{rcodeboxtable}{R output of \texttt{TailClassifier()} for the power-decaying random sample with seed 2023, sample size 4,000, and parameter 1.5.}{r-box-Pareto0.5}
\begin{lstlisting}
$Conclusion
[1] "The data suggest a power decaying tail."
$CI
[1] "95% CI based on information from v_left = 20 to v_right = 54"
    Tail_Type          CI_level    lwr         upr         PointEstimate
1   Power              0.95        1.4815060   1.6778960   1.4978049
2   Sub-Exponential    0.95        0.2806986   0.3260212   0.3015819
3   Near-Exponential   0.95        2.5527540   3.1643117   2.8596888
\end{lstlisting}
\end{rcodeboxtable}

\subsubsection*{\textcolor{black}{Illustrative Example: Sub-Exponential Distribution}}

The sub-exponential simulation is generated from the continuous density
\[
g(x)=\frac{e}{4}e^{-\sqrt{x}}, \qquad x>1.
\]
A continuous sample drawn from \(g\) is converted into a discrete sample by taking the floor of each observation. By Theorem~\ref{thm-bin_tail_type_same}, the resulting discrete distribution preserves the sub-exponential tail type, with parameter \(0.5\).

\begin{figure}[htbp]
    \centering
    \includegraphics[width=\linewidth]{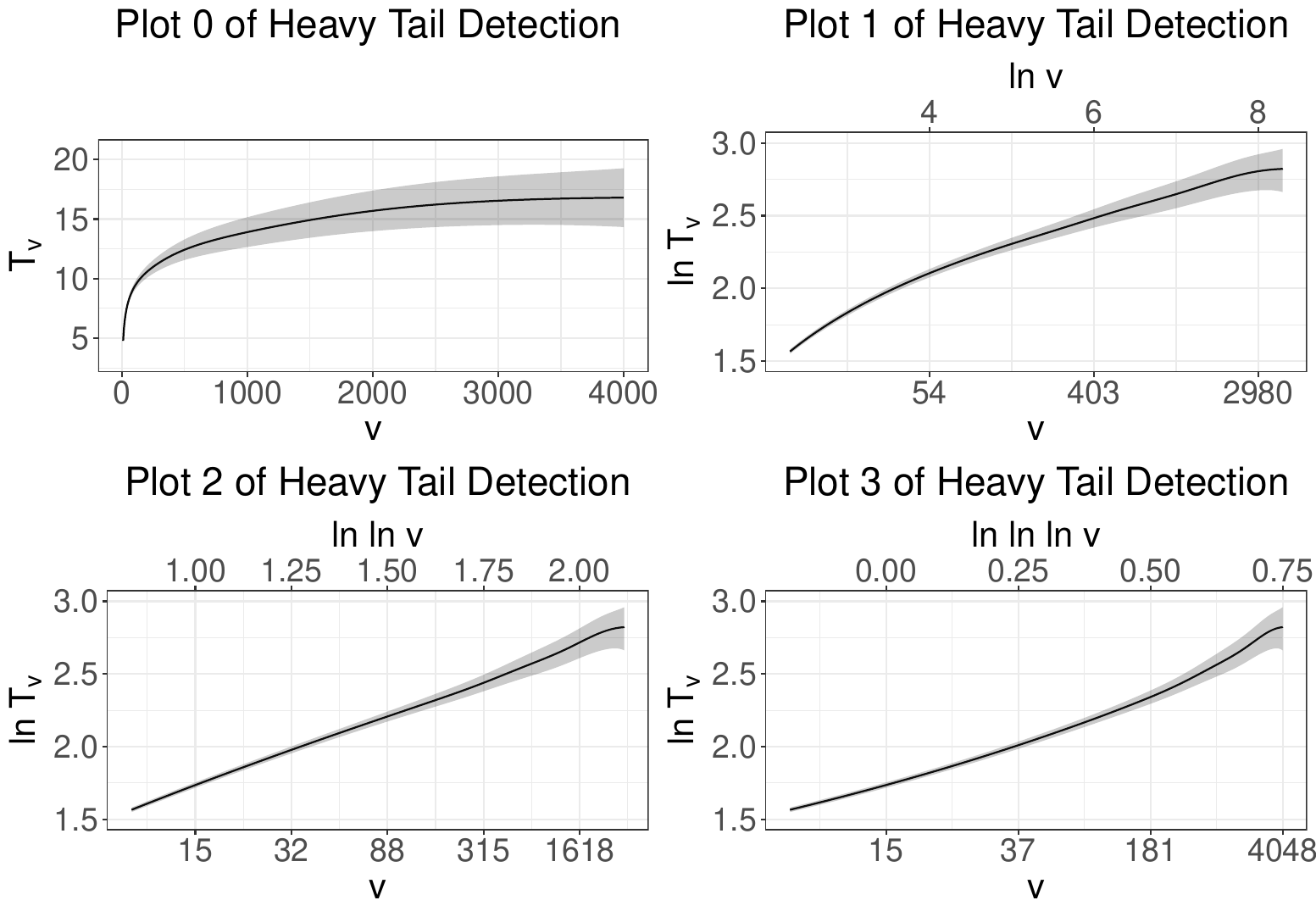}
    \caption{Entropic plots for a \textcolor{black}{single} random sample of size \(10,000\) generated from a sub-exponentially decaying tail distribution with parameter \(0.5\).}
    \label{fig-subexp0.5}
\end{figure}

A discrete random sample of size \(10,000\) was generated using the same procedure with random seed \(2023\). A larger sample size is used here because the sub-exponential tail is thinner than the power tail, making tail information more difficult to observe. The corresponding entropic plots, along with 95\% confidence bands, are shown in Figure~\ref{fig-subexp0.5}. In Plot~0, the upward trend indicates a tail thicker than exponential. A slight downturn after approximately \(v=3,500\) suggests that very large values of \(v\) are unreliable for this sample, as discussed in Section~\ref{sec-how-to-select-tail-profile}. Among Plots~1, 2, and 3, Plot~2 exhibits the clearest linear trend, supporting a sub-exponential tail. For parameter estimation, a tail profile from \(v=\floor{\exp(e^{1.25})}=32\) to \(v=\floor{\exp(e^{1.5})}=88\) was selected. The output is reported in Table~\ref{r-box-subexp0.5}, where the 95\% confidence interval \((0.4665,0.5905)\) contains the true value \(0.5\).

\begin{rcodeboxtable}{R output of \texttt{TailClassifier()} for the sub-exponential decaying random sample with seed 2023, sample size 10,000, and parameter 0.5.}{r-box-subexp0.5}
\begin{lstlisting}
$Conclusion
[1] "The data suggest a sub-exponential decaying tail."
$CI
[1] "95% CI based on information from v_left = 32 to v_right = 88"
    Tail_Type          CI_level    lwr         upr         PointEstimate
1   Power              0.95        4.1158794   5.696927    4.2947332
2   Sub-Exponential    0.95        0.4664778   0.590545    0.5209081
3   Near-Exponential   0.95        0.9477974   1.563450    1.2572453
\end{lstlisting}
\end{rcodeboxtable}

\subsubsection*{\textcolor{black}{Illustrative Example: Near-Exponential Distribution}}

\begin{figure}[htbp]
    \centering
    \includegraphics[width=\linewidth]{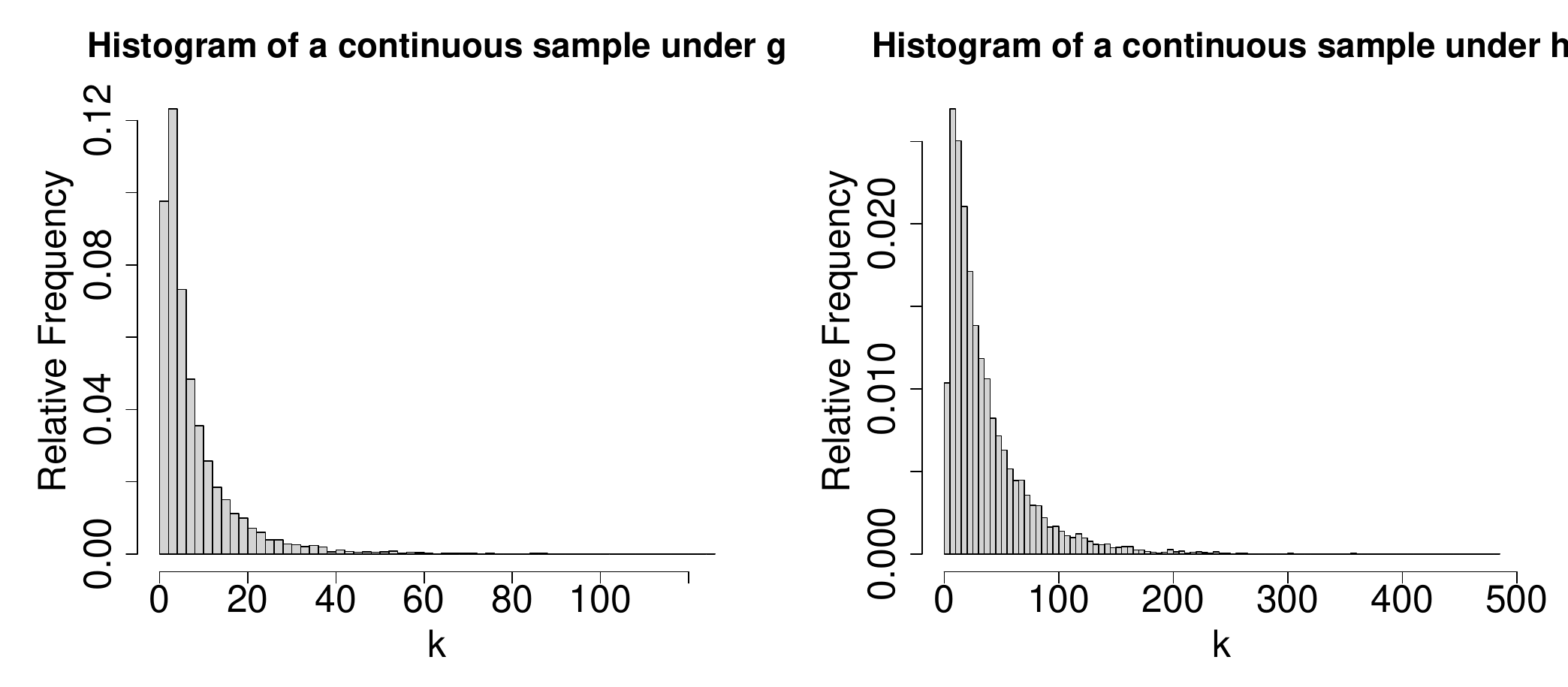}
    \caption{Relative-frequency histograms for two continuous random samples of size \(100,000\) generated from the sub-exponential density \(g\) and the near-exponential density \(h\).}
    \label{fig-hist}
\end{figure}

The near-exponential simulation is generated from the continuous density
\[
h(x)=c\exp\left(-\frac{x}{\ln^2 x}\right), \qquad x>1,
\]
where \(c\approx 1/5.69797\) is the normalizing constant. A continuous sample drawn from \(h\) is converted into a discrete sample by taking the floor of each observation. By Theorem~\ref{thm-bin_tail_type_same}, the resulting discrete distribution preserves the near-exponential tail type, with parameter \(2\). Figure~\ref{fig-hist} shows relative-frequency histograms for continuous samples of size \(100,000\) drawn from \(g\) and \(h\), illustrating that the near-exponential behavior becomes apparent only in the distant tail.

\begin{figure}[htbp]
    \centering
    \includegraphics[width=\linewidth]{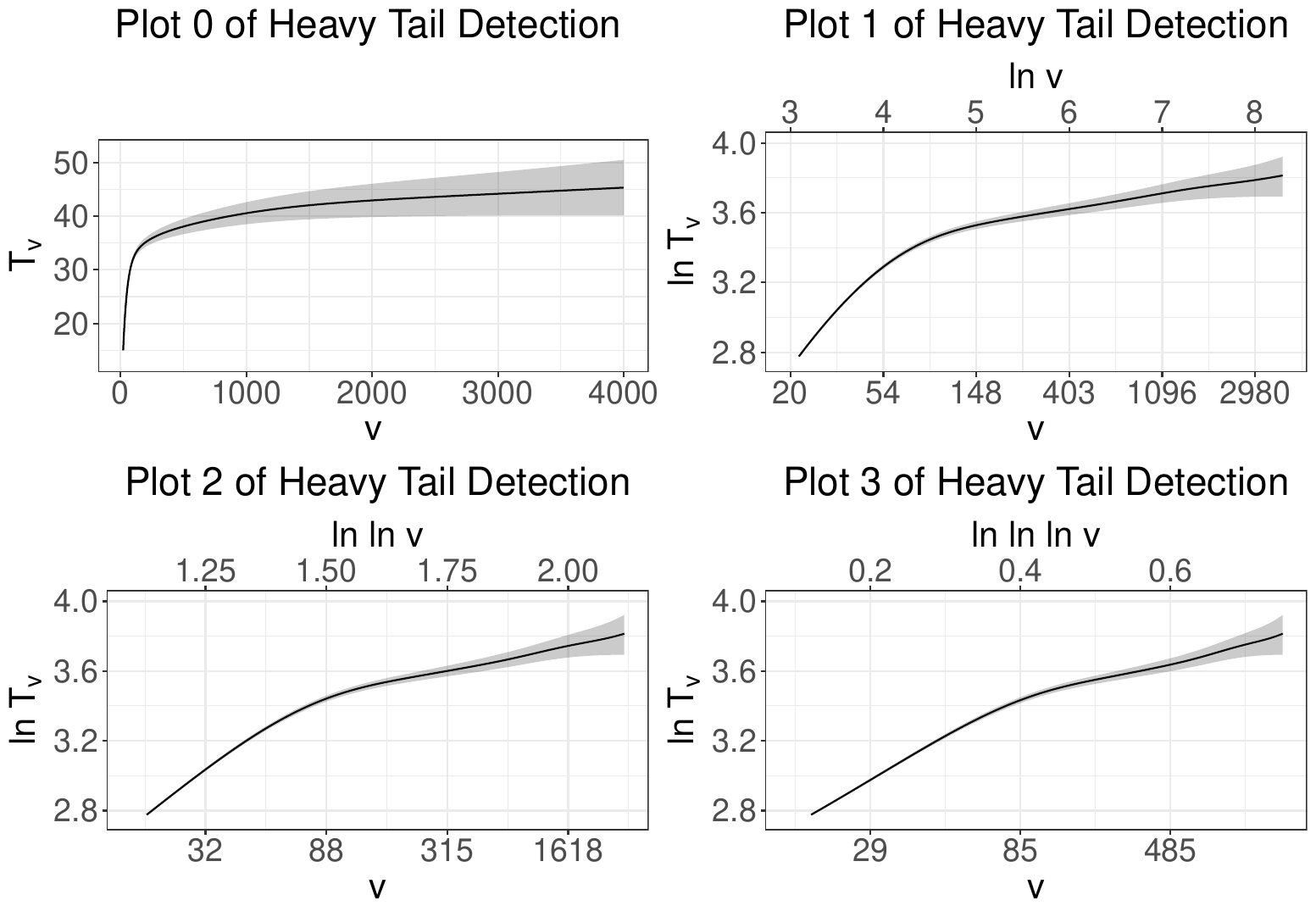}
    \caption{Entropic plots for a \textcolor{black}{single} random sample of size \(10,000\) generated from a near-exponentially decaying tail distribution with parameter \(2\).}
    \label{fig-nearexp_1_2}
\end{figure}

A discrete random sample of size \(10,000\) was generated from \(h\) with random seed \(2023\). The corresponding entropic plots are shown in Figure~\ref{fig-nearexp_1_2}. In Plot~0, the upward trend indicates a tail thicker than exponential. The behavior of Plots~1, 2, and 3 suggests that values \(v>315\) are unreliable for this sample; see Section~\ref{sec-how-to-select-tail-profile}. Restricting attention to \(v\leq 315\), Plot~3 exhibits the clearest linear trend, supporting a near-exponential tail. For parameter estimation, a tail profile from \(v=\floor{\exp(\exp(e^{0.3}))}=47\) to \(v=\floor{\exp(\exp(e^{0.4}))}=85\) was selected. The output is reported in Table~\ref{r-box-nearexp_1_2}, where the 95\% confidence interval \((1.7770,2.3415)\) contains the true value \(2\).

\begin{rcodeboxtable}{R output of \texttt{TailClassifier()} for the near-exponential decaying random sample with seed 2023, sample size 10,000, and parameter 2.}{r-box-nearexp_1_2}
\begin{lstlisting}
$Conclusion
[1] "The data suggest a near-exponential decaying tail."
$CI
[1] "95% CI based on information from v_left = 47 to v_right = 85"
    Tail_Type          CI_level    lwr         upr         PointEstimate
1   Power              0.95        2.7832439   3.3041243   2.850258
2   Sub-Exponential    0.95        0.3772524   0.4438964   0.407786
3   Near-Exponential   0.95        1.7770380   2.3415456   2.060008
\end{lstlisting}
\end{rcodeboxtable}

\subsubsection*{\textcolor{black}{Monte Carlo Results}}

For each of the three distributions, \(1,000\) additional random samples were generated to assess classification accuracy and interval performance. Although the tail profile could be selected separately for each sample, a common tail profile was used within each simulation setting for simplicity: \(v=20\) to \(v=54\) for the power distribution, \(v=32\) to \(v=88\) for the sub-exponential distribution, and \(v=47\) to \(v=85\) for the near-exponential distribution. The results are summarized in Table~\ref{table-simu-results}.

\begin{table}[htbp]
\caption{\textcolor{black}{Monte Carlo} results based on \(1,000\) replications for each distribution. Percentages \((a\%,b\%,c\%)\) in parentheses under the correct classification rate indicate the percentages classified as power, sub-exponential, and near-exponential, respectively. Numbers in parentheses in the last column are the true parameter values.}
\label{table-simu-results}
\centering
\begin{tabular}{cccccc}
    \toprule
     &  & \textbf{Correct} & \textbf{95\%} & \textbf{Mean} & \textbf{Mean} \\
    \textbf{Type} & \textbf{n} & \textbf{Classification} & \textbf{CI} & \textbf{CI} & \textbf{Estimated} \\
     &  & \textbf{Rate} & \textbf{Coverage} & \textbf{Width} & \textbf{Parameter} \\
        \midrule
    Power    & 4,000  & 100\% (100\%, 0\%, 0\%)        & 96.4\% & 0.196 & 1.4824 (1.5) \\
    Sub-Exp  & 10,000 & 79.3\% (0.4\%, 79.3\%, 20.3\%) & 100\%  & 0.115 & 0.501 (0.5) \\
    Near-Exp & 10,000 & 100\% (0\%, 0\%, 100\%)        & 100\%  & 0.567 & 2.0501 (2) \\
    \bottomrule
\end{tabular}
\end{table}

The \textcolor{black}{Monte Carlo} results show that TENT performs well for the power and near-exponential settings considered here, with correct classification rates of \(100\%\). The sub-exponential setting is more challenging: although the parameter estimates and interval coverage remain accurate, approximately \(20.3\%\) of the samples are classified as near-exponential. This is consistent with the fact that sub-exponential and near-exponential tails can be difficult to distinguish over finite sample ranges, especially when the effective tail profile is limited.

\section{Discussion on Tail-Profile Selection}
\label{sec-how-to-select-tail-profile}

Selecting an appropriate tail profile is important for the practical use of TENT. The role of the index range \(v_1\leq v\leq v_2\) is analogous to the choice of a threshold, or equivalently the number or sample fraction of upper order statistics, in classical tail-index estimation. Hill's estimator \cite{hill1975simple}, for example, has motivated substantial work on threshold and sample-fraction selection, including adaptive procedures and regression-based diagnostics; see, for example, \cite{hall1985adaptive, beirlant1996tail, drees1998selecting}. Similar issues arise here: choosing \(v\) too small may cause the profile to reflect the body of the distribution rather than the tail, whereas choosing \(v\) too large may lead to unstable estimates because the relevant low-probability symbols are rarely observed. The following guidelines summarize practical diagnostics used in the numerical examples. A systematic theory of optimal tail-profile selection is left for future work.

A lower bound on \(v\) is needed to ensure that the asymptotic approximations used for parameter estimation are meaningful. In particular, to ensure that \(\ln x_v>1\) in Lemma~\ref{lemmaNEsolution} of Appendix~\ref{subsec-proof-main}, the selected range should satisfy
\[
v>\exp\{\max(\hat{\theta},e)\},
\]
where \(\hat{\theta}\) denotes the point estimate for the selected tail type. The upper end of the range is guided by the empirical entropic plots. Unusual behavior at large \(v\), such as a downturn, an inflection point, or rapidly widening confidence bands, may indicate that the corresponding part of the empirical profile is unreliable. Excessively large values of \(v\) may also create numerical instability and increase estimation variability. In the numerical examples considered here, values of \(v\) up to approximately \(3,000\) were computationally feasible in R 4.4.2 without \texttt{Rmpfr} \cite{r-Rmpfr}, while the simulation studies in Section~\ref{subsec-simu} suggest that substantially smaller ranges, often with \(v\leq 100\), may already be sufficient.

The shape of the entropic plots provides additional diagnostic information. When Plot~0 shows an upward trend, indicating a tail thicker than exponential, the transformed plots should display behavior consistent with one of the benchmark regimes. In particular, the plot corresponding to the selected regime should exhibit an approximately linear trend over the chosen range. The remaining plots should have shapes compatible with the ordering of the transformations implied by Theorem~\ref{th2}. For example, if Plot~1 is approximately linear, then the evidence favors a power tail; if Plot~2 is approximately linear, the evidence favors a sub-exponential tail; and if Plot~3 is approximately linear, the evidence favors a near-exponential tail. Sectional downturns, inflection points, or inconsistent curvature patterns may indicate that the chosen range includes unstable parts of the empirical profile.

Confidence intervals also provide useful diagnostics. If the confidence interval for the selected tail type has a lower bound equal to \(0\), an infinite upper bound, or, in the sub-exponential case, an upper bound reaching \(1\), then the data may not contain enough information to support reliable parameter estimation over the chosen range. Similarly, if the point estimate lies outside its reported confidence interval, the selected profile should be treated with caution. Such behavior does not necessarily invalidate the classifier, but it suggests that the chosen range of \(v\) should be adjusted or that the sample contains limited information about the tail.

In summary, the following criteria are used as practical guidelines for selecting a tail profile:
\begin{enumerate}[nosep, label=(\arabic*)]
    \item The selected range should satisfy \(v>\exp\{\max(\hat{\theta},e)\}\), where \(\hat{\theta}\) is the point estimate for the selected tail type.
    \item The selected plot among Plots~1--3 should exhibit an approximately linear trend over the chosen range.
    \item If Plot~0 trends upward, then the selected range should avoid sectional downturns or inflection points in Plots~1--3.
    \item Plot~1 should not exhibit clear convexity over the selected range, since this would suggest behavior thicker than the power-law benchmark.
    \item If Plot~1 is approximately linear, then Plots~2 and~3 should exhibit compatible convex curvature.
    \item If Plot~2 is approximately linear, then Plot~1 should exhibit compatible concave curvature and Plot~3 compatible convex curvature.
    \item If Plot~3 is approximately linear, then Plots~1 and~2 should exhibit compatible concave curvature.
    \item The lower bound of the confidence interval for the selected tail type should be positive.
    \item For power and near-exponential tails, the upper confidence bound should be finite.
    \item For sub-exponential tails, the upper confidence bound should be less than \(1\).
    \item The point estimate should lie within its corresponding confidence interval.
\end{enumerate}

If several of these diagnostics fail, the tail-profile range should be reconsidered. Persistent failure across reasonable choices of \(v_1\) and \(v_2\) may indicate that the sample does not contain sufficient tail information to distinguish reliably among the candidate regimes.

\section{Conclusion}
\label{sec-conclusion}

This article introduces the Turing--Entropic Tail Classifier (TENT), a nonparametric approach to tail inference based on domain-of-attraction theory on countable alphabets. The method uses Turing-type and entropic quantities to construct a tail profile for the ranked probability mass function. By comparing the behavior of the empirical tail profile with theoretical benchmarks, TENT distinguishes among exponential-type, near-exponential, sub-exponential, and power-law ranked tails. In the power-law case, the resulting inference is closely related in spirit to Hill-type tail-index estimation, while the broader tail-profile framework avoids imposing a power-law assumption at the outset.

For heavier-than-exponential tails, TENT provides point and interval estimates for selected tail parameters. Although the method is developed primarily for countable alphabets and discrete distributions, Theorem~\ref{thm-bin_tail_type_same} in Appendix~\ref{subsec-proof-bin} enables application to continuous observations through equal-width discretization. The numerical examples illustrate the workflow of the classifier and show that tail-profile selection plays an important role in finite-sample performance. A systematic theory for optimal tail-profile selection is left for future work.

The method is implemented in the R package \texttt{TailClassifier}, available on CRAN. It is intended as a preliminary diagnostic tool for identifying plausible tail regimes and guiding subsequent parametric modeling or domain-specific analysis.

\backmatter

\section*{Acknowledgements}

An earlier version of this article is available as a preprint on arXiv \cite{zhang2022entropicb}.

\section*{Declarations}

\begin{itemize}
\item Funding: Not applicable.
\item Competing interests: The authors declare that they have no competing interests.
\item Ethics approval and consent to participate: Not applicable.
\item Consent for publication: Not applicable.
\item Data availability: The data used in the numerical examples are described in the article. Additional data availability information can be provided upon reasonable request.
\item Materials availability: Not applicable.
\item Code availability: The method is implemented in the R package \texttt{TailClassifier}, available on CRAN.
\item Author contributions: Both authors contributed to all aspects of the work.
\end{itemize}

\bigskip

\begin{appendices}

\section{Theoretical Support}
\label{sec-proofs}

\subsection{Supporting Results for Entropic Plots}
\label{subsec-proof-main}

\textcolor{black}{The main theoretical justification for the proposed classifier consists of the consistency result in Theorem~\ref{th1} and the population divergence rates in \eqref{powerpi}--\eqref{nearexppi} established in Theorem~\ref{th2}. Theorem~\ref{th1} is proved first because its consistency argument is self-contained and does not require the auxiliary results used to derive the population divergence rates. The supporting lemmas are then introduced before the proof of Theorem~\ref{th2} for the three representative tail forms in \eqref{dist1}--\eqref{dist3}.}

\begin{proof}[Proof of Theorem \ref{th1}]
        Noting that
        \[
        \frac{n^{v+1}[n-(v+1)]!}{n!}
        =
        \frac{1}{\prod_{j=0}^{v}(1-j/n)},
        \]
        $T_v$ may be expressed as
        \[
        T_v
        =
        \frac{v}{1-v/n}
        \sum_{k=1}^{\infty}
        \hat p_k(n)
        \prod_{j=0}^{v-1}
        \frac{1-\hat p_k(n)-j/n}{1-j/n}.
        \]
        For fixed \(v\), the prefactor \((1-v/n)^{-1}\) converges to \(1\).

		For each fixed $k$, by the Strong Law of Large Numbers: 
		\(
		\hat{p}_k(n) \;\xrightarrow[n\to\infty]{\text{a.s.}}\; p_k.
		\)
		Hence, for each $j = 0,1,\dots,v-1$, 
		\[
		\frac{\,1 - \hat{p}_k(n) - j/n\,}{\,1 - j/n\,}
		\;\xrightarrow[n\to\infty]{a.s.}\;
		1 - p_k.
		\]
		It follows that
		\[
		\hat{p}_k(n)\,
		\prod_{j=0}^{v-1}
		\frac{\,1 - \hat{p}_k(n) - j/n\,}{\,1 - j/n\,}
		\;\xrightarrow[n\to\infty]{\text{a.s.}}\;
		p_k\,\bigl(1 - p_k\bigr)^v \;
		\]
            \text{for each fixed $k$ and $v$.} Next, note that
            \[
            0 \le \hat{p}_k(n) \prod_{j=0}^{v-1}\frac{1 - \hat{p}_k(n) - j/n}{1 - j/n}\le \hat{p}_k(n)
            \]
            \text{and}
            \[
            \quad
            \sum_{k=1}^\infty \hat{p}_k(n) = 1.
            \]
            Since
            \[
            \max_{k \geq 1}\left|\hat{p}_k - p_k\right|\xrightarrow{a.s.}0 \text{ (Lemma 2 in \cite{zhang2023several})}, 
            \]
            it follows by the General Lebesgue Dominated Convergence Theorem (Theorem 19 in \cite{royden2010real}) that
            \[
            \{\hat{p}_k(n)\}\xrightarrow[n\to\infty]{a.s.}\{p_k\},
            \]
            thus allowing passage of the limit inside the infinite sum:
		\[
		\sum_{k=1}^\infty
		\hat{p}_k(n)
		\prod_{j=0}^{v-1}
		\frac{\,1 - \hat{p}_k(n) - j/n\,}{\,1 - j/n\,}
		\xrightarrow[n\to\infty]{\text{a.s.}}
		\sum_{k=1}^\infty
		p_k\,(1-p_k)^v.
		\]
		It follows that
		\[
		T_v
		\; \;\xrightarrow[n\to\infty]{\text{a.s.}}\;
		v\sum_{k=1}^\infty p_k\,(1-p_k)^v
		\;=\;\tau_v.
		\]
		Since $\tau_v > 0$, taking the ratio on both sides gives
        \[
        T_v / \tau_v \xrightarrow[n\to\infty]{\text{a.s.}} 1
        \]
        for each fixed $v$.
\end{proof}

Next, the convergence rates, (\ref{powerpi}), (\ref{subexppi}) and (\ref{nearexppi}), are established for the three distributions in (\ref{dist1}), (\ref{dist2}) and (\ref{dist3}) respectively. Toward that end, a general result established by \cite{molchanov2018entropic} is re-stated below as Lemma \ref{molchanov}.

Consider the following five conditions on the underlying distribution $\mathbf{p}=\{p_{k};k\geq 1\}$.
\begin{enumerate}[nosep]
	\item[C1:] Assume $p_{k+1}/p_{k}\rightarrow 1$;
	\item[C2:] assume there exists a smooth $C^2(R_{+})$ interpolation $p(x)$ for $x>0$ such that $p_{k}=p(k)$ for all $k$, $p(0)<\infty$, and $p'(x)<0$ for $x\geq x_{0}$ where $x_{0}$ is a sufficiently large number;
	\item[C3:] assume the underlying interpolation $p(x)$ satisfies 
	\begin{enumerate}[nosep]
		\item $(\ln p(x))'=p'(x)/p(x)\nearrow 0$, as $x\rightarrow \infty$ and
		\item $\lim_{x\rightarrow \infty}p^2(x)/p'(x)=0$. 
	\end{enumerate}
	\item[C4:] assume $(p'(x)/p(x))'\geq 0$; and 
	\item[C5:] assume that there exists a constant $\gamma \in [0,1]$, such that $$\limsup_{x\rightarrow \infty}\frac{p''(x)p(x)-(p'(x))^2}{(p'(x))^2}=1-\gamma.$$
\end{enumerate}
\begin{lemma}
	Under Conditions C1-C5, 
    \[
    \tau_{v}\asymp p(x_{v})/|p'(x_{v})|,
    \]
    where $x_{v}$ is the root of $vp(x)=1$.
	\label{molchanov}
\end{lemma}

To prove Theorem \ref{th2}, Lemma \ref{lemmaNEsolution} below is needed.
\begin{lemma} Let $x_{v}$ be the root of equation
	\beq
	\frac{x^{1/\beta}}{\ln x}=(\ln v)^{1/\beta}
	\label{ex_rootxofxZ}
	\eeq where $\beta>0$ is a positive constant. As $v\rightarrow \infty$, $\ln x_{v}\asymp \ln \ln v$ and $x_{v}\asymp (\ln v)( \ln \ln v)^{\beta}$.
	\label{lemmaNEsolution}
\end{lemma}

\begin{proof}[Proof of Lemma \ref{lemmaNEsolution}]
Condition (\ref{ex_rootxofxZ}) can be rewritten as $x_v^{1/\beta} = (\ln x_v)(\ln v)^{1/\beta}$. For large \(x_v\) (and hence large \(v\)), \(\ln x_v > 1\), so $x_v^{1/\beta} > (\ln v)^{1/\beta} \Rightarrow \ln x_v > \ln\ln v$. On the other hand, for sufficiently large \(x_v\), $x_v^{1/(2\beta)} < {x_v^{1/\beta}}/{\ln x_v} = (\ln v)^{1/\beta}$, which implies $\ln x_v < 2\ln\ln v$. Thus, $\ln\ln v < \ln x_v < 2\ln\ln v$, i.e., $\ln x_v \asymp \ln\ln v$.
Multiplying by \((\ln v)^{1/\beta}\) yields $(\ln v)^{1/\beta}\ln\ln v < x_v^{1/\beta} < 2(\ln v)^{1/\beta}\ln\ln v$. Raising both sides to the power \(\beta\) gives $(\ln v)(\ln\ln v)^\beta < x_v < 2^\beta (\ln v)(\ln\ln v)^\beta$, that is $x_v \asymp (\ln v)(\ln\ln v)^\beta$.
\end{proof}

\begin{proof}[Proof of Theorem \ref{th2}]
	For each of the representative tail forms in 
	(\ref{dist1}), (\ref{dist2}) and (\ref{dist3}), \cite{molchanov2018entropic} showed that Conditions C1-C5 are all satisfied. It only remains to derive the rates of divergence using Lemma \ref{molchanov}, that is, $p(x_{v})/|p'(x_{v})|$. 
	
	For (\ref{dist1}), the root of the equation $p(x)=cx^{-\lambda}=1/v$ is
	$x_{v}=(cv)^{1/\lambda}$, and $p'(x)=-\lambda cx^{-(1+\lambda)}$. It follows that $p(x_{v})=1/v$ and $p'(x_{v})=-\lambda c^{-(1/\lambda)} v^{-(1+1/\lambda)}$ and that $p(x_{v})/|p'(x_{v})|=(c^{1/\lambda}/\lambda)v^{1/\lambda}$.
	
	For (\ref{dist2}), the root of the equation $p(x)=ce^{-\lambda x^{\alpha}}=1/v$ is
    \[
    x_v=\left(\frac{\ln(cv)}{\lambda}\right)^{1/\alpha},
    \]
    and
    \[
    |p'(x_v)|
    =
    \lambda\alpha x_v^{\alpha-1}p(x_v)
    =
    \frac{\lambda\alpha}{v}
    \left(\frac{\ln(cv)}{\lambda}\right)^{(\alpha-1)/\alpha}.
    \]
    Therefore
    \[
    \frac{p(x_v)}{|p'(x_v)|}
    =
    \frac{1}{\lambda\alpha}
    \left(\frac{\ln(cv)}{\lambda}\right)^{(1-\alpha)/\alpha}
    \asymp
    (\ln v)^{1/\alpha-1}.
    \]
    
	For (\ref{dist3}), the root of the equation $p(x)=ce^{-\lambda x/(\ln x)^{\beta}}=1/v$ does not have an analytic form. However, rewriting the equation gives
    \[
    \frac{x^{1/\beta}}{\ln x}
    =
    \left(\frac{\ln(cv)}{\lambda}\right)^{1/\beta}.
    \]
    Applying Lemma \ref{lemmaNEsolution}, the root of the equation satisfies 
	\[\ln x_{v}\asymp \ln \ln \left((cv)^{1/\lambda}\right)=-\ln \lambda+\ln \ln (cv). \]
	Noting 
	\[p'(x)=-p(x)\lambda \frac{(\ln x)^{\beta}-\beta (\ln x)^{\beta-1}}{(\ln x)^{2\beta}},
	\]it follows that 
	\[ \frac{p(x_{v})}{|p'(x_{v})|}=\frac{(\ln x_{v})^{\beta+1}}{\ln x_{v}-\beta}\sim (\ln x_{v})^{\beta}\asymp (\ln \ln v)^{\beta}.
	\] 
\end{proof}

\subsection{Confidence Bands}
\label{subsec-proof-CI}
Next, theoretical support for the interval estimates is provided. \textcolor{black}{For every pair of positive integers $(u, v)$, let
	\begin{align*}
	Z_{u,v} =& \frac{[n - (u + v)]!}{n!} n^{u+v} \sum_{k \geq 1} \left\{ 1 \left[\hat{p}_k \geq \frac{u}{n}\right]\right. \\
	&\left.\left[ \prod_{i=0}^{u-1} \left( \hat{p}_k - \frac{i}{n} \right) \right] 
	\left[ \prod_{j=0}^{v-1} \left( 1 - \hat{p}_k - \frac{j}{n} \right) \right] 
	\right\},
	\end{align*}
	where $1 \left[\hat{p}_k \geq \frac{u}{n}\right]$ is an indicator function that takes value 1 if $\hat{p}_k \geq \frac{u}{n}$, and 0 otherwise.}

\begin{lemma}[Corollary 2.1 in Ref.~\cite{zhang2016statistical}]
Let 
\textcolor{black}{\[ 
\hat{\sigma}_{Z_v} = \left\{ Z_{2v} -2vZ_{2,2v-1} +v^2Z_{3,2v-2} -v^2\left(\frac{1}{v}Z_v-Z_{2,v-1}\right)^2 \right\}^{1/2}. 
\] }
Then 
\[ 
\frac{\sqrt{n}(Z_v-\zeta_v)} {\hat{\sigma}_{Z_v}} \cond N(0,1). 
\] 
\label{lemma-zv-normal} 
\end{lemma} 

\begin{lemma} 
\[ 
\frac{\sqrt{n}(T_v-\tau_v)} {\hat{\sigma}_{T_v}} \cond N(0,1), 
\] 
where 
\textcolor{black}{\[ 
\hat{\sigma}_{T_v} = v\left\{ Z_{2v} -2vZ_{2,2v-1} +v^2Z_{3,2v-2} -v^2\left(\frac{1}{v}Z_v-Z_{2,v-1}\right)^2 \right\}^{1/2}. 
\]}
\label{lemma-Tv-normal} 
\end{lemma}

Lemma \ref{lemma-Tv-normal} is an immediate result of Lemma \ref{lemma-zv-normal}, as $T_v = vZ_v$ and $\tau_v = v\zeta_{v}$ by definition. Lemma \ref{lemma-Tv-normal} offers a confidence band for selected tail profile. From (\ref{powerpi}), (\ref{subexppi}) and (\ref{nearexppi}), one may obtain the relationship between the slope of the linear line in the selected plot and the parameter of interest. For example, for power tail distribution, $\lambda \asymp 1/m$, where m is the slope that $ m = \Delta \ln \tau_v / \Delta \ln v$. Similarly, for sub-exponential tail distribution, $\alpha \asymp 1/(m+1)$, where $m = \Delta \ln \tau_v / \Delta \ln \ln v$. And for near-exponential tail distribution, $\beta \asymp m = \Delta \ln \tau_v / \Delta \ln \ln \ln v$. Therefore, obtaining confidence inference for $\lambda$, $\alpha$, and $\beta$, is essential to obtain confidence intervals for the slope in each situation.

Toward that end, let the $100(1-a)\%$ confidence band of $\tau_v$ given by Lemma \ref{lemma-Tv-normal} be 
$$
C_{a} = \{(l_{v, a}, u_{v, a}): v \text{ in tail profile, from } v_1 \text{ to } v_2\}.
$$
A conservative $100(1-a)\%$ confidence interval for the slope within the selected tail profile may be conservatively obtained as 
$$
\text{lower}_{m,a} = \frac{\ln l_{v_2,a} - \ln u_{v_1,a}}{\Delta x}
$$
and 
$$
\text{upper}_{m,a} = \frac{\ln u_{v_2,a} - \ln l_{v_1,a}}{\Delta x},
$$
where
$$
\Delta x = \begin{cases}
	\ln v_2 - \ln v_1, & \text{Power tail} \\
	\ln \ln v_2 - \ln \ln v_1, &  \text{Sub-Exponential tail} \\
	\ln \ln \ln v_2 - \ln \ln \ln v_1, &  \text{Near-Exponential tail}
\end{cases}.
$$

Given the domains are $\lambda > 1$, $\alpha \in (0,1)$, and $\beta > 0$, $100(1-a)\%$ confidence intervals for them are established as

{\footnotesize
\[
\begin{cases}
	\lambda: & \left( \max \left\{ 1, \nicefrac{1}{\text{upper}_{m,a}} \right\}, 
	\nicefrac{1}{\max\left\{0, \text{lower}_{m,a}\right\}} \right) \\
	\alpha: & \left( \max \left\{ 0, \nicefrac{1}{ ( \max\{0, \text{upper}_{m,a}\} + 1} ) \right\}, 
	\min \left\{ 1, \nicefrac{1}{\text{lower}_{m,a} + 1} \right\} \right) \\
	\beta: & \left( \max \left\{ 0, \text{lower}_{m,a} \right\}, 
	\max\left( 0, \text{upper}_{m,a} \right) \right)
\end{cases},
\]
}
where $1/0 =: \infty$.

The confidence bands are based on the fixed-\(v\) asymptotic normal approximation as \(n\to\infty\). The parameter intervals above are conservative transformations of endpoint confidence bands over the selected tail profile. Their finite-sample reliability therefore depends on the choice of \(v_1\) and \(v_2\), as discussed in Section~\ref{sec-how-to-select-tail-profile}. The selected \(v\) should satisfy \(2v+1\le n\), so that all terms appearing in \(\hat\sigma_{T_v}\) are well-defined.

\subsection{Binning Certain Continuous Variable with Equal Width Preserves Tail Information}
\label{subsec-proof-bin}

\begin{theorem}
Let \(X\) have density \(f\) on \([x_0,\infty)\), and fix a bin width \(c>0\). Define
\[
q_m=\int_{x_m}^{x_m+c}f(t)\,dt,
\qquad x_m=x_0+mc.
\]
If
\[
q_m \asymp f(x_m)
\qquad\text{as }m\to\infty,
\]
then equal-width binning preserves the tail decay rate of the density, up to multiplicative constants. In particular, the ranked tail class is unchanged for the benchmark families considered in this article.
\label{thm-bin_tail_type_same}
\end{theorem}

\begin{proof}[Proof of Theorem~\ref{thm-bin_tail_type_same}]
The binned distribution assigns probability \(q_m\) to the \(m\)th tail bin. By assumption, \(q_m\asymp f(x_m)\), so the binned probability sequence has the same asymptotic decay rate as the original density evaluated along the bin grid. Multiplicative constants do not affect the tail classes considered \textcolor{black}{in this article}.
\end{proof}

For \(f(x)\propto x^{-\lambda}\), fixed-width integration gives
\[
q_m=\int_{x_m}^{x_m+c}t^{-\lambda}\,dt\asymp x_m^{-\lambda}.
\]

For \(f(x)\propto \exp(-\lambda x^\alpha)\), \(\alpha\in(0,1)\),
\[
q_m\sim c\exp(-\lambda x_m^\alpha),
\]
because \(f(x_m+t)/f(x_m)\to 1\) uniformly for \(0\le t\le c\).

For \(f(x)\propto \exp\{-\lambda x/(\log x)^\beta\}\),
\[
q_m\sim c\exp\left\{-\frac{\lambda x_m}{(\log x_m)^\beta}\right\},
\]
by the same fixed-width local flatness argument.

\subsection*{\textcolor{black}{Sensitivity Analysis}}

As a sensitivity check, the AMZN example in Section~\ref{sec-amazon} is reanalyzed using a coarser common bin width of \(10^{-3}\), rather than \(10^{-4}\). The results are reported in Tables~\ref{r-box-l3} and~\ref{r-box-r3}.

\begin{rcodeboxtable}{R output of \texttt{TailClassifier()} for left-tail data with bin width \(\delta=10^{-3}\).}{r-box-l3}
\begin{lstlisting}
$Conclusion
[1] "The data suggest a near-exponential decaying tail."
$CI
[1] "95% CI based on information from v_left = 29 to v_right = 85"
    Tail_Type          CI_level    lwr         upr         PointEstimate
1   Power              0.95        3.6936025   11.27970    4.1828236
2   Sub-Exponential    0.95        0.4015830   0.74405     0.5187681
3   Near-Exponential   0.95        0.4636841   2.00862     1.2504039
\end{lstlisting}
\end{rcodeboxtable}

\begin{rcodeboxtable}{R output of \texttt{TailClassifier()} for right-tail data with bin width \(\delta=10^{-3}\).}{r-box-r3}
\begin{lstlisting}
$Conclusion
[1] "The data suggest a near-exponential decaying tail."
$CI
[1] "95% CI based on information from v_left = 33 to v_right = 90"
    Tail_Type          CI_level    lwr         upr         PointEstimate
1   Power              0.95        3.8482823   12.3699742  4.2917731
2   Sub-Exponential    0.95        0.3976930   0.7567093   0.5190283
3   Near-Exponential   0.95        0.4417704   2.0809903   1.2732939
\end{lstlisting}
\end{rcodeboxtable}

Comparisons between Tables \ref{r-box-l4} and \ref{r-box-l3}, as well as Tables \ref{r-box-r4} and \ref{r-box-r3}, show that both discretization widths lead to the same near-exponential classification for the left and right tails. The point estimates and confidence intervals vary with the bin width, as expected, but the qualitative tail classification remains unchanged. In particular, the near-exponential estimates obtained under one bin width fall within the corresponding confidence intervals obtained under the other bin width.

\end{appendices}


\bibliography{references_db}

\end{document}